\documentclass{article}

\usepackage{arxiv}

\usepackage[utf8]{inputenc} % allow utf-8 input

\usepackage[T1]{fontenc}    % use 8-bit T1 fonts
\usepackage{hyperref}       % hyperlinks
\usepackage{url}            % simple URL typesetting
\usepackage{booktabs}       % professional-quality tables
\usepackage{amsfonts}       % blackboard math symbols
\usepackage{nicefrac}       % compact symbols for 1/2, etc.
\usepackage{microtype}      % microtypography
\usepackage{lipsum}
\usepackage{graphicx}
\usepackage{multirow}
\usepackage{amsmath}
\usepackage[table]{xcolor}
\usepackage{float}
\usepackage{caption}
\usepackage{graphicx}
\usepackage{tabularx}

\usepackage{array}

\usepackage{subcaption}
\usepackage{pdfpages}

\usepackage{booktabs}
\usepackage{array}    
\usepackage{makecell} 
\graphicspath{ {./images/} }

\title{Multimodal Integration Challenges in Emotionally Expressive Child Avatars for Training Applications}

\author{
 Pegah Salehi  \\
  SimulaMet \\
  Oslo, Norway\\
  \texttt{pegah@simula.no} \\
   \And
 Sajad Amouei Sheshkal  \\
  SimulaMet \\
  Oslo, Norway\\
  \texttt{sajad.amouei@gmail.com} \\
  \And
 Vajira Thambawita  \\
  SimulaMet \\
  Oslo, Norway\\
  \texttt{vajira@simula.no} \\
  \And
  Michael A. Riegler  \\
  Simula \\
  Oslo, Norway\\
  \texttt{michael@simula.no} \\
  \And 
 Pål Halvorsen \\
  SimulaMet \\
  Oslo, Norway\\
  \texttt{paalh@simula.no} \\
}

\begin{document}
\maketitle
\begin{abstract}

 Dynamic facial emotion is essential for believable AI-generated avatars; however, most systems remain visually inert, limiting their utility in simulations such as virtual training for investigative interviews with abused children. We introduce and evaluate a real-time architecture fusing Unreal Engine 5 MetaHuman rendering with NVIDIA Omniverse Audio2Face to translate vocal prosody into high-fidelity facial expressions on photorealistic child avatars. Due to limitations in synthetic voice options, both avatars were voiced using a young adult female TTS model, selected from two different systems in an attempt to better match each character. This compromise introduces a confounding factor. Voice-age mismatches and prosodic or emotional differences may disrupt audiovisual alignment. We implemented a two-PC setup decoupling language and speech synthesis from GPU-intensive rendering, designed to support low-latency interaction in desktop and VR environments. A between-subjects study ($N=70$) using audio+visual and visual-only conditions assessed perceptual impacts as participants rated emotional clarity, facial realism, and empathy for two avatars expressing joy, sadness, and anger. 

Results show that while avatars could express emotions recognizably, particularly sadness and joy, recognition of anger dropped markedly without audio, highlighting the role of vocal cues in conveying high-arousal states. Interestingly, silencing the clips improved perceived realism by removing mismatches between facial animation and voice, especially where age or emotional tone were incongruent. These findings highlight that perceived believability hinges on the interplay between audiovisual congruence and facial geometry: a mismatched voice can undermine even well-crafted expressions, while a congruent one can enhance weaker visuals. This trade-off presents an ongoing challenge for designing emotionally coherent avatars in sensitive training contexts.

\end{abstract}

% Keywords

%%%%%%%%%%%%%%%%%%%%%%%%%%%%%%%%%%%%%%%%%%

\section{Introduction}

AI-generated avatars are rapidly becoming integral components of digital interaction across diverse domains, with applications spanning virtual, augmented, and mixed reality (VR/AR/XR) platforms, virtual telepresence, embodied digital assistants, e-commerce, entertainment, education, and potential therapeutic interventions. The efficacy and adoption of these avatars critically depend on their perceived believability and their capacity to engage users in naturalistic interactions. For an avatar to "feel alive" and effectively bridge the human-computer interaction gap, it must convincingly replicate key facets of human social signaling~\cite{danvevcek2025supervising}. Among these, the dynamic expression of facial emotions stands out as particularly vital. Facial expressions serve as a primary, often subconscious, channel for non-verbal communication, conveying a wealth of affective information that shapes interpersonal understanding, relationship dynamics, and the interpretation of intent~\cite{pei2024affective}. Consequently, an avatar's ability to display appropriate, recognizable, and dynamically evolving facial emotions is paramount for creating engaging, trustworthy, and effective virtual agents. This paper investigates the technical feasibility of creating such avatars and evaluates their perceptual impact, particularly in training scenarios.

perceived believability hinges on the interplay between audiovisual congruence and facial geometry

\begin{figure}
\centering
\includegraphics[scale=0.60, trim=20 0 23 0, clip]{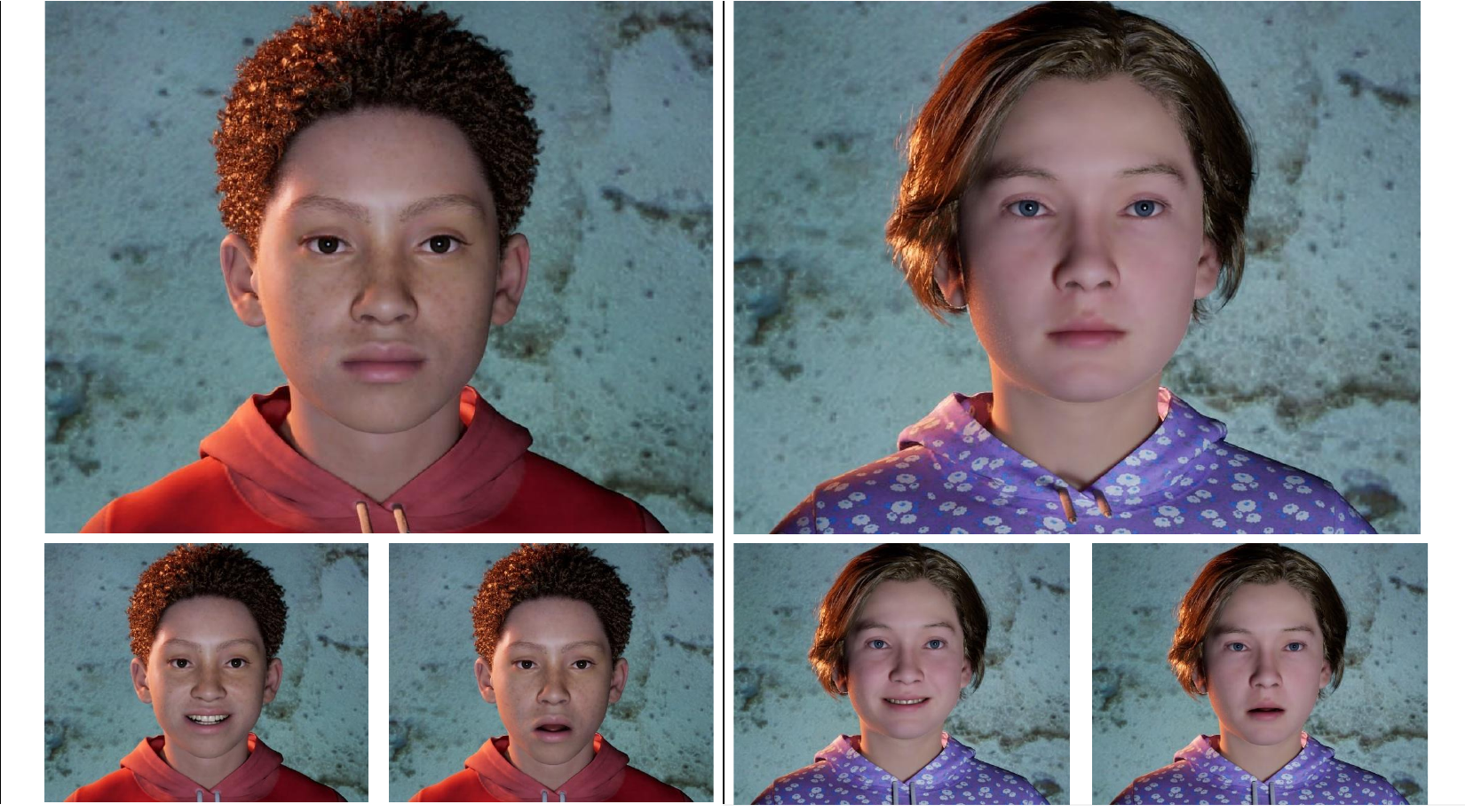}
\caption{Example frames of the two avatars used in the user study: "Emory" (male-presenting) and "Amelia" (female-presenting). The top row displays their neutral facial expressions. The bottom rows show expressions of joy (left) and sadness (right) for each avatar, as used in the study. For clarity, only two emotions are visualized here; the full set of expressions, including anger, is available in the supplementary videos provided in the project’s GitHub repository~\cite{salehi2025emotionally}.}
\label{fig:metahuman}
\end{figure}

While the importance of such affective realism is broadly recognized, its consistent and effective implementation is foundational and challenging in high-stakes domains. One such domain is simulation-based training for professionals who conduct forensic interviews with children, often concerning experiences of abuse or trauma. In this context, previous virtual-child systems have primarily focused on dialogue flow and question typology. However, empirical evaluations have repeatedly highlighted two significant shortcomings~\cite{salehi2024theoretical}: firstly, the avatars themselves are frequently emotionally inert, their faces remaining neutral regardless of the conversational context, thereby offering no micro-expressions that might signal anxiety, relief, or distress. Secondly, practitioners often find the overall visual fidelity of these avatars inadequate, with subtle inconsistencies in proportions, shading, or motion disrupting immersion and undermining trust in the simulation. Collectively, these deficits impede core learning objectives; interviewers are unable to accurately judge whether a question soothes or unsettles the virtual child, cannot adequately fine-tune their empathic responses, and consequently struggle to transfer refined interviewing skills to live cases where non-verbal cues are indispensable. Furthermore, the capacity to render these digital characters in real time is essential for such interactive applications, facilitating immediate responsiveness and a heightened sense of presence.

Addressing these limitations requires a focused research endeavor into developing AI avatar systems capable of rich, dynamic facial emotion expression with real-time responsiveness. This research directly tackles this objective by introducing and assessing a novel, distributed architecture engineered for high-fidelity emotional expressiveness. This system represents a significant advancement by integrating the sophisticated rendering capabilities of Unreal Engine 5 (UE5) MetaHuman software~\cite{pan2022research} with the intelligent, audio-driven facial animation provided by NVIDIA's Omniverse Audio2Face~\cite{nvidia_audio2face_3d}. The system’s distributed architecture employs two synchronized computational nodes to effectively balance processing demands by decoupling language understanding and speech generation from GPU-intensive rendering, thereby enabling low-latency interaction. Figure~\ref{fig:metahuman} displays representative screenshots of the two photorealistic MetaHuman avatars, "Amelia" and "Emory," utilized in this study.

In summary, this study offers the following contributions:

\begin{itemize}
    \item Development of a modular, real-time pipeline integrating Unreal Engine MetaHuman and NVIDIA Audio2Face to generate emotionally expressive child avatars for interactive applications.
    
    \item Empirical investigation ($N=70$) comparing audio+visual and visual-only conditions, revealing how mismatched modalities can reduce perceived realism, and in some cases, visual-only presentations may be preferred.
    
    \item Identification of a perceptual asymmetry across emotional arousal levels: while low-arousal emotions (e.g., sadness, joy) were recognizable from facial cues alone, high-arousal emotions (e.g., anger) were more dependent on audiovisual congruence.
    
    \item Observation that avatar facial morphology (e.g., soft vs. angular features) influences emotion interpretation, suggesting the need for morphology-emotion alignment in sensitive contexts such as empathy training.
    
    \item Extraction of practical design lessons from participant feedback, including the impact of audiovisual desynchrony, the trade-off between expression clarity and naturalism, and the importance of age-appropriate voice synthesis.
\end{itemize}

\section{Related Work}

The pursuit of emotionally expressive AI avatars lies at the confluence of affective computing, human-computer interaction (HCI), and advancements in multimodal systems, all striving to endow machines with a semblance of human emotional intelligence~\cite{aranha2019adapting, maas2011learning, tian2001recognizing, pereira2018empirical}. While foundational work in affective computing centered on the detection and modeling of emotional cues, the emphasis has progressively shifted towards the dynamic and adaptive expression of these emotions in avatars to foster user engagement and perceived believability. This aligns with broader HCI principles where anthropomorphism in digital agents consistently correlates with enhanced user engagement, perceived credibility, and overall satisfaction~\cite{mcduff2018designing, lalwani2018implementation, nass2000machines, jack2015human}. The significance of expressive capabilities is underscored by research in applied domains like virtual therapy and education, where emotionally inert or insufficiently expressive avatars can lead to misunderstandings, diminished rapport, and ultimately, reduced system effectiveness~\cite{cabanac2002emotion, schachter1962cognitive, gratch2004evaluating, gratch2005lessons}.

This need for dynamic emotional expressiveness becomes particularly acute in high-stakes interaction scenarios, such as the virtual training environments for child forensic interviews. In these contexts, an avatar's ability to display contextually appropriate and recognizable emotions can profoundly influence user perception and training outcomes. However, existing virtual-child systems tailored for such training have predominantly concentrated on dialogue management and question typologies. Empirical evaluations of these systems have consistently identified critical shortcomings: specifically, avatars often lack dynamic facial affect, remaining emotionally static, and their overall visual fidelity is frequently judged as inadequate~\cite{salehi2024theoretical}. These deficiencies impair training efficacy, as trainees cannot accurately gauge a child's non-verbal reactions to their questioning or refine their empathic responses, which are skills paramount in real-world interviews.

The technological evolution towards more expressive avatars has been considerable. Initial efforts often involved 2D video-based avatars that, while capable of displaying emotions, tended to produce rigid outputs through direct mappings from input signals~\cite{ji2021audio, papantoniou2022neural, thies2020neural}. The advent of 3D parametric models, such as FLAME~\cite{li2017learning}, offered more granular control over facial geometry and pose. Subsequently, sophisticated speech-driven animation systems like EMOTE~\cite{danvevcek2023emotional} and EmoTalk~\cite{peng2023emotalk} have emerged, aiming to disentangle and map speech content and prosody to facial expressions. However, as these systems push the boundaries of realism, they often encounter challenges in achieving both high-fidelity emotional rendering and the stringent real-time performance demanded by interactive, immersive simulations. Many existing solutions struggle to seamlessly integrate state-of-the-art rendering quality with low-latency, audio-driven emotional animation, particularly for photorealistic child avatars where subtlety and believability are crucial.

Collectively, the literature highlights a persistent gap: the need for AI avatar systems that not only generate rich and dynamic facial emotions but do so with high visual fidelity and in real-time, especially for demanding applications like child interview simulations. While advancements in speech-driven animation and realistic rendering are notable, their integration into a cohesive, responsive system that meets these multifaceted requirements remains a significant hurdle. Our research directly confronts this challenge by proposing and evaluating a novel, distributed architecture. This system uniquely fuses the high-fidelity rendering capabilities of Unreal Engine 5's MetaHuman with NVIDIA Omniverse Audio2Face for real-time, prosody-driven facial emotion on photorealistic child avatars. By decoupling language processing and speech generation from GPU-intensive rendering via a two-PC setup, our approach is specifically designed to achieve the low-latency interaction essential for effective and immersive training simulations.

\begin{figure}
\centering
\includegraphics[scale=0.92]{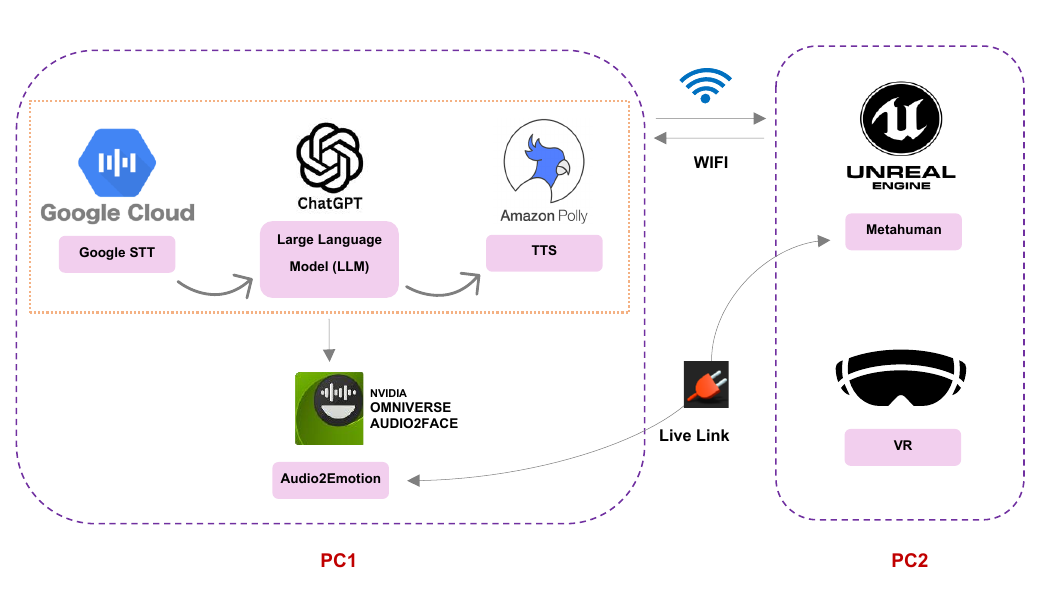}
\caption{Architecture for real-time emotionally expressive child avatars. PC1 processes speech input (Google Cloud STT), generates a text response via a large language model (LLM), and synthesizes speech using Amazon Polly TTS. The resulting audio is analyzed by NVIDIA Audio2Face to extract emotional parameters for facial animation. PC2 renders a customizable Metahuman avatar (Unreal Engine) driven by this emotional data in real time. Synchronization is achieved using the Live Link plugin, and the system supports deployment in both desktop and VR environments.}
\label{fig:arch5}
\end{figure}

\section{Methodology}

\subsection{System Design and Implementation}

To overcome the limitations of previous systems~\cite{salehi2024theoretical} for emotional expression in AI avatars, this research introduces a novel architecture that integrates Unreal Engine 5 Metahuman~\cite{pan2022research} for high-fidelity facial rendering with NVIDIA’s Omniverse Audio2Face~\cite{nvidia_audio2face_3d} for generating emotionally expressive facial animations from audio input. The source code and implementation details for the system described in this study are publicly available~\cite{salehi2025emotionally}. Below, we provide a brief overview of these two core platforms:

\begin{itemize}
  \item \textbf{Unreal Engine 5 MetaHuman} integrates advanced real-time graphics technologies such as Nanite geometry, the Lumen lighting system, and optimized post-processing to deliver film-quality visuals without compromising interactivity~\cite{EpicGames_MetaHuman, EpicGames_RigLogic_WhitePaper}. Using the MetaHuman Creator, researchers can import head scans or digital sculpts and generate fully rigged characters that are customizable in every anatomical detail and exportable at multiple levels of detail (LOD) for deployment across platforms ranging from mobile devices to virtual reality.

  The system includes a preconfigured facial rig built on UE5’s Control Rig, which offers hundreds of easily retargetable controls. This allows expressions to be driven by live capture, audio input, or keyframe animation. To achieve photorealistic skin, which is essential for avoiding the uncanny valley, MetaHuman incorporates subsurface scattering (SSS)~\cite{EpicGames_SSS_UE5}, a rendering method that simulates the way light penetrates and diffuses through real skin. Additional realism is provided through groom-based hair and physically accurate materials. Together, these features enable a fast end-to-end pipeline for producing responsive and emotionally expressive avatars designed for AI-driven training scenarios involving child interactions.

  \item \textbf{NVIDIA Omniverse Audio2Face} is the core component responsible for generating lifelike facial animations in response to audio input. This generative AI tool analyzes speech to automatically produce synchronized lip movements and expressive facial animations in real time, thereby enhancing the emotional realism of virtual characters. It includes an intuitive character retargeting pipeline that enables users to apply animations to custom 3D characters, including seamless integration with Unreal Engine MetaHumans.

  The system provides a Blendshape Generation widget that converts audio into facial expressions using templates such as NV 46 Pose or ARKit 51 Pose. Although the tool primarily supports full-face animation, it can also transfer movements to individual facial features such as the eyes, jaw, and tongue.

  A key advantage of Audio2Face is its real-time performance, which makes it particularly well-suited for interactive applications including virtual reality, digital assistants, and serious games. It also supports traditional animation workflows via live playback or baked outputs. Additional features include multilingual input support, simultaneous animation of multiple characters, and manual adjustment of emotion intensity using facial action unit (FAU) controls. The interface also allows users to fine-tune automatic emotion extraction from audio using settings such as emotion detection range, strength, contrast, smoothing, and the number of concurrently active emotions. These capabilities enable fine-grained, context-aware emotional expression, which is especially important in sensitive domains such as child interview simulations.

  In this study, the “exaggerated” mode in Audio2Face was used to enhance the clarity and distinctiveness of facial expressions in short-duration clips, facilitating consistent emotion recognition among participants. This setting was chosen based on preliminary observations during development, as the standard and automatic modes often produced expressions that were too subtle to be reliably perceived within the limited duration of the stimuli.

\end{itemize}

The proposed architecture, illustrated in Figure~\ref{fig:arch5}, enables real-time interactivity through the coordinated operation of two synchronized computational nodes. PC1 handles all speech and language processing. The user's spoken input is first transcribed using the Google Cloud Speech-to-Text (STT) engine~\cite{Google2023_SpeechToText}, which leverages advanced models such as Google’s foundation model Chirp to provide high accuracy across diverse accents and acoustic environments, with support for multilingual input. The resulting transcription is processed by OpenAI’s ChatGPT Large Language Model (LLM)~\cite{OpenAI2024_ChatGPT}, which generates a context-appropriate textual response. This response is then synthesized into speech using Amazon Polly’s Text-to-Speech (TTS) system~\cite{AWS2025_PollyNeuralVoices}, which offers lifelike multilingual voices and allows customization of intonation, pronunciation, and speaking style.

In parallel, NVIDIA Audio2Face analyzes the prosodic and spectral features of the synthesized speech~\cite{Daly2024_Audio2Face}. It extracts frame-level emotional parameters, such as indicators of joy, fear, or sadness, based on vocal characteristics. These parameters are typically represented as blendshape weights or facial action unit (FAU) activations and are transmitted via Wi-Fi to PC2, the rendering node.

PC2 is responsible for avatar rendering and real-time facial animation. It runs Unreal Engine 5’s Metahuman framework, which produces high-fidelity child avatars with anatomically accurate facial rigs. The emotional parameters received from PC1 are used to drive the avatar’s expressions in real time by mapping the blendshape weights or FAU activations to the MetaHuman Control Rig via the Live Link plugin~\cite{NVIDIA2025_Audio2FaceLiveLink}. This ensures low-latency synchronization between facial animation and the synthesized voice.

This distributed system design separates the computationally intensive processes of natural language understanding and speech generation from the demanding task of real-time, high-resolution avatar rendering. While this architecture is designed for responsiveness, potential sources of audiovisual desynchrony can still arise from factors such as network latency or processing delays within either computational node. Nevertheless, this separation enables the system to maintain both overall responsiveness and visual realism, supporting immersive and emotionally expressive interactions in applications such as child interview training.

\subsection{User Study}

To examine the feasibility of incorporating emotional expressiveness into AI-driven avatars through dynamic facial emotion expression and to assess its perceptual impact, we conducted a user study focused on three key dimensions: emotional clarity, facial realism, and empathic connection. These dimensions were chosen as they are key indicators of avatar believability and effectiveness in interactive simulations. Two photorealistic MetaHuman avatars, named Emory and Amelia, were used to assess perceived emotional expressiveness under different conditions. Representative screenshots of the avatars are shown in Figure~\ref{fig:metahuman}.

The study followed a between-subjects design with two experimental conditions: one condition included both voice and facial expression (audio+visual condition), while the other featured facial expression only (visual-only condition). This design was informed by prior findings suggesting that audio and visual modalities contribute differently to emotion perception, specifically that audio cues are more salient for detecting changes in emotional arousal, while visual cues are more informative for identifying emotional valence~\cite{wu2021multimodal}.

A total of 77 participants were initially recruited via the Prolific platform. Although initial eligibility was filtered using Prolific’s pre-screening tools (e.g., age, language fluency, and device requirements), an additional quality control step was applied during data cleaning. Based on clearly invalid or inconsistent responses (e.g., repeated ratings across stimuli or contradictory answers), 7 participants were excluded from further analysis. The final sample included 70 participants whose responses met the criteria for attentiveness and internal consistency. Participants were randomly assigned to one of two experimental conditions ($N = 35$ per condition) using Prolific’s built-in randomization tool. To minimize order effects, we manually randomized the sequence of avatar stimuli within each condition, varying both avatar identity and emotional expression. All participants in a given condition viewed the same randomized stimulus order. The sample exhibited diversity in gender, age, academic/professional background, and familiarity with virtual environments.

The participant pool ($N=70$) was evenly distributed by gender, comprising 35 women (50.0\%) and 35 men (50.0\%). The predominant age group was 25--34 years (40 participants, 57.1\%), followed by 35--44 years (16 participants, 22.9\%). Smaller segments included participants aged 18--24 years (10 participants, 14.3\%) and 45--54 years (4 participants, 5.7\%). The academic and professional backgrounds of the participants were highly varied, with numerous unique responses. 
%Among these, the most frequently cited fields were IT and Finance, each reported by 3 participants (4.3\%).

Regarding familiarity with virtual environments and avatars, a significant portion of participants reported interacting "Occasionally" (32 participants, 45.7\%). Other interaction frequencies included "Rarely" (15 participants, 21.4\%), "Frequently" (12 participants, 17.1\%), and "Very frequently" (9 participants, 12.9\%). A small fraction (2 participants, 2.9\%) reported "Never" interacting with avatars. This demographic profile indicates a sample primarily composed of young to middle-aged adults from diverse professional domains, possessing a general, though not necessarily expert-level, familiarity with avatar interactions.

Participants viewed a series of six high-definition video clips, the presentation order of which was counterbalanced to avoid order effects. In these clips, the avatars expressed three emotional states: sadness, anger, and joy (see Figure~\ref{fig:emotion-scripts}). These specific emotions were selected as they represent a range of valence and arousal levels and are relevant emotions that children might express in sensitive interview contexts. In the audio+visual condition, speech was synthesized using two different TTS systems: OpenAI’s TTS\footnote{\url{https://ttsopenai.com/}} for the \textit{Emory} avatar and NaturalReaders\footnote{\url{https://www.naturalreaders.com/}} for the \textit{Amelia} avatar. We evaluated several available text-to-speech options and selected these two voices based on their perceived suitability, as judged by two of the researchers. Although child-like voices were technically available, a young adult female voice was ultimately preferred for both avatars due to its better perceptual match with the visual and emotional characteristics of the characters. This choice was supported by findings from our prior studies~\cite{salehi2022more}, which showed that female voices can be effectively used to represent child characters in training scenarios, regardless of gender. Additionally, we found the child-like voices to be overly juvenile in tone and pitch, which created a mismatch with the avatars’ appearance and demeanor, both of which were designed to resemble pre-teens or young adolescents rather than very young children.

\begin{figure}[ht]
\centering
\renewcommand{\arraystretch}{1.6} % Adds vertical spacing
\begin{tabular}{|p{1.5cm}|p{12cm}|}
\hline
\textbf{Sadness} & \textit{"I thought today would be different... but it wasn’t. No one talked to me at school. I just sat alone… again. I don’t know why they don’t like me. Maybe I’m just not good enough."} \\
\hline
\textbf{Anger} & \textit{"It’s not fair! I told them it wasn’t my fault, but no one listened. They just blamed me like always! I’m so tired of this. I try to do the right thing, but it never matters!"} \\
\hline
\textbf{Joy} & \textit{"Today was amazing! My teacher said I did a great job on my homework, and my friends invited me to play during lunch. Everything just felt really good ... like things are finally going my way!"} \\
\hline
\end{tabular}
\caption{Scripted utterances used for emotional avatar expressions in the user study.}
\label{fig:emotion-scripts}
\end{figure}

\begin{figure}
\centering
\includegraphics[scale=0.2]{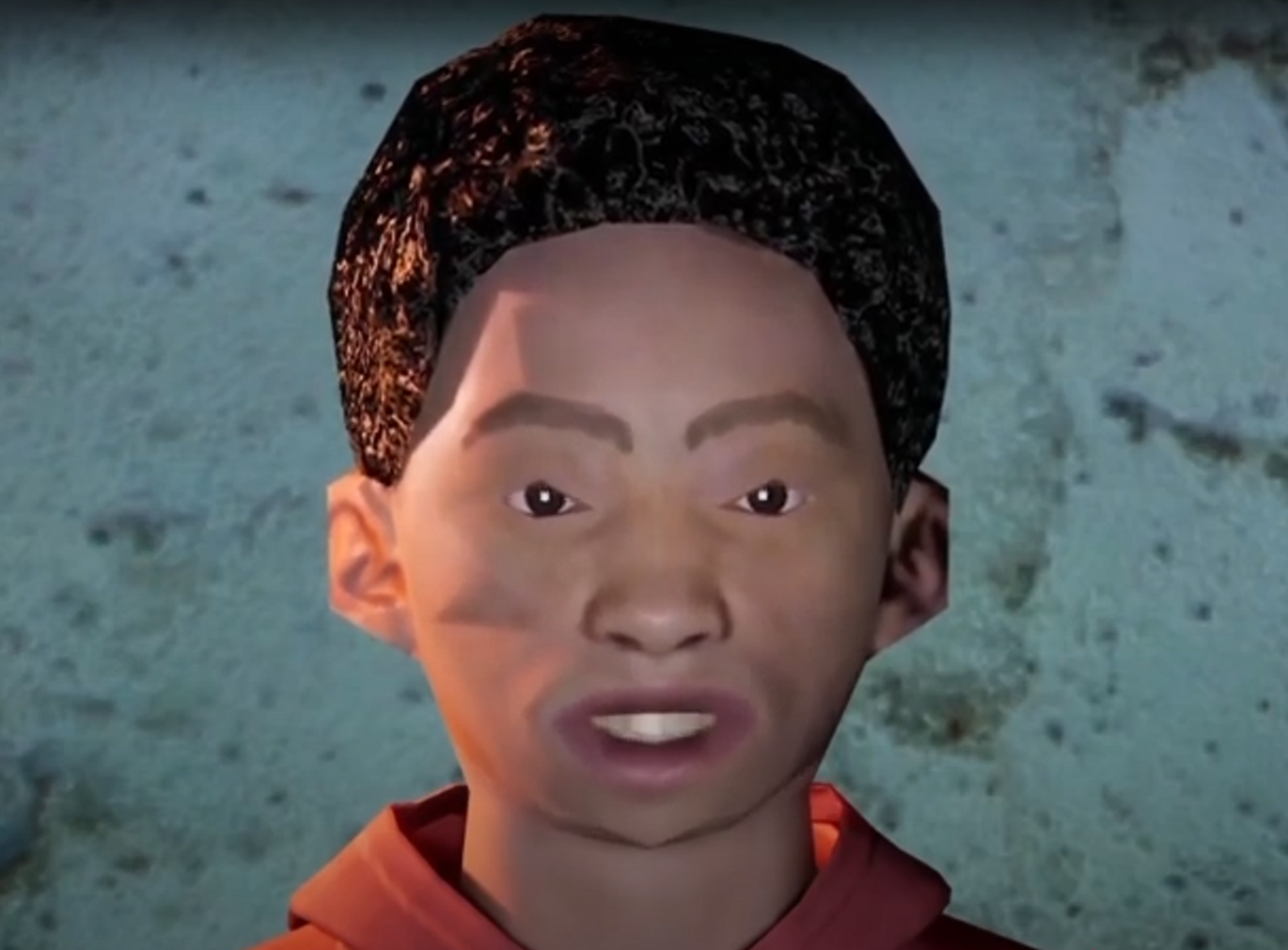}
\caption{Low-fidelity anchoring stimulus used at the beginning of the user study to calibrate participants’ expectations regarding avatar realism. }
\label{fig:anchoring}
\end{figure}

The six video clips were generated using the described system architecture. For the audio+visual condition, the scripts were synthesized into audio, and this audio was processed by NVIDIA Audio2Face to generate facial animation data. This animation data was then applied to the MetaHuman avatars in Unreal Engine 5 and rendered into video clips. For the visual-only condition, the same facial animation data was applied to the avatars and rendered into video clips without the synthesized audio track.

To manage participant expectations and establish a perceptual anchor, we began with a deliberately low-fidelity stimulus, following evidence that anchoring effects require the presence of both semantic and numerical cues~\cite{onuki2021stimuli}. The inclusion of this initial low-fidelity stimulus was intended to provide a baseline or reference point for participants, potentially making their subsequent ratings of the higher-fidelity main stimuli more calibrated and less susceptible to being skewed by initial unfamiliarity or overly high expectations for avatar realism. An example frame from this anchoring stimulus is shown in Figure~\ref{fig:anchoring}. By first exposing them to a clearly less sophisticated avatar, we aimed to help participants better contextualize and evaluate the nuances of the emotionally expressive avatars that formed the core of the study. 

After viewing the main stimuli, participants completed a questionnaire comprising three main sections (see Table~\ref{tab:questionnaire-summary}). items were rated on a five-point Likert scale. Emotion recognition and perceived facial realism were assessed for each individual video, while perceived realism across different contributing aspects (e.g., facial expressions, voice tone) and overall empathy toward the avatars were measured in a final, post-video section of the questionnaire. The final section included an open-ended question inviting participants to reflect on what stood out most about the avatars. 

\begin{table}[ht]
\centering
\scriptsize
\caption{Structure and content of the user study questionnaire, including measures of perceived realism, emotional expressiveness, and overall impressions. The full questionnaire is available in the project’s GitHub repository~\cite{salehi2025emotionally}.}
\renewcommand{\arraystretch}{1.5} 
\begin{tabular}{m{3.5cm}m{2.5cm}m{6.9cm}}
\hline
\textbf{Section} & \textbf{Question Type} & \textbf{Content} \\
\hline
\textbf{Emotion Recognition} & Likert scale & Extent to which the avatar expressed: Anger, Sadness, Joy, Fear, Disgust \\
\hline
\textbf{Facial Realism} & Likert scale & Realism of facial expressions in each video \\
\hline
\textbf{Realism Contributors} & Likert scale & Contribution of each factor to perceived realism: \newline - Facial expressions \newline - Visual appearance \newline - Voice tone* \newline - Dialogue content* \\
\hline
\textbf{Empathy/Connection} & Likert scale & Emotional connection or empathy felt toward the avatars overall \\
\hline
\textbf{Open-ended Reflection} & Free text & One sentence describing what stood out most about the avatars \\
\hline
\textbf{Demographics} & Multiple choice & Gender, age group, academic/professional background, frequency of avatar interaction \\
\hline
\end{tabular}
\begin{flushleft}
\footnotesize{\textit{* Voice tone and dialogue content were excluded from the visual-only condition.}}
\end{flushleft}
\label{tab:questionnaire-summary}

\end{table}

\subsection{Ethical Considerations}

This study involved the evaluation of emotional expressiveness in \textit{synthetic} avatars by adult participants via anonymous online surveys. All visual and auditory stimuli were fully computer-generated and did not include any real individuals or personal data. This study did not require formal ethical approval, as it involved no interaction with vulnerable groups, no collection of sensitive personal data, and no deception. Participants were recruited through the Prolific platform and were informed about the nature and duration of the study prior to participation. No formal consent form was presented beyond Prolific’s standard participation agreement.
While the study featured simulated emotional expressions (e.g., sadness and anger) in photorealistic child avatars, no content warnings were provided prior to exposure. Some participants reported feeling uncomfortable or uneasy, particularly in response to expressions of sadness or anger. This is noted as a potential ethical consideration. Future implementations should consider including content advisories to reduce the risk of discomfort for sensitive viewers.
%According to the guidelines of the Norwegian National Research Ethics Committees (NREC) and Simula Research Laboratory, This study did not require formal ethical

\section{Results and Analysis} \label{sec:result}

To comprehensively assess how participants interpreted the emotional expressions of the avatars, we conducted both quantitative and qualitative analyses. 

\subsection{Quantitative Analysis}

The quantitative analysis in this study focused on how participants perceived and interpreted the emotional expressions of AI-generated child avatars across different experimental conditions. Specifically, we examined three main response categories:

\begin{enumerate}
    \item \textbf{Emotion recognition}, defined as how accurately participants identified the intended emotional expression (anger, sadness, joy) after each video,
    \item \textbf{Perceived facial expression realism}, referring to the extent to which facial animations appeared realistic in each video, and
    \item \textbf{Perceived realism across contributing aspects} (e.g., facial expressions, visual appearance, voice tone, dialogue content), along with the \textbf{empathic connection} participants felt toward the avatars, assessed in a post-exposure questionnaire.
\end{enumerate}

We first computed descriptive statistics (mean and standard deviation) for each category. To assess effects of condition, avatar identity, and emotion scenario on these outcomes, we employed cumulative link mixed-effects models (CLMMs) using the \texttt{ordinal} package in \textsf{R}. These models incorporated fixed effects for \textit{Condition} (audio+visual vs. visual-only), \textit{Avatar} (Amelia vs. Emory), and \textit{Scenario} (anger, sadness, joy), along with their interaction terms. Random intercepts were included for participants and, where applicable, for emotion or video stimuli, to account for within-subject and item-level variability. Model fitting was performed using a logit link and adaptive Gauss–Hermite quadrature (\texttt{nAGQ = 10}) to ensure robust convergence and estimation accuracy.

Before examining each response category in detail, we begin by validating the robustness and goodness-of-fit of our statistical models. The results are then presented across the following dimensions: emotion recognition, perceived facial realism, and perceived realism/empathy. For completeness and cross-method validation, supplementary t-test analyses are reported in Appendix Table~\ref{tab:main_t-test} and Table~\ref{tab:t-test-avatar}.

\begin{figure}
\centering
\includegraphics[scale=0.42, clip]{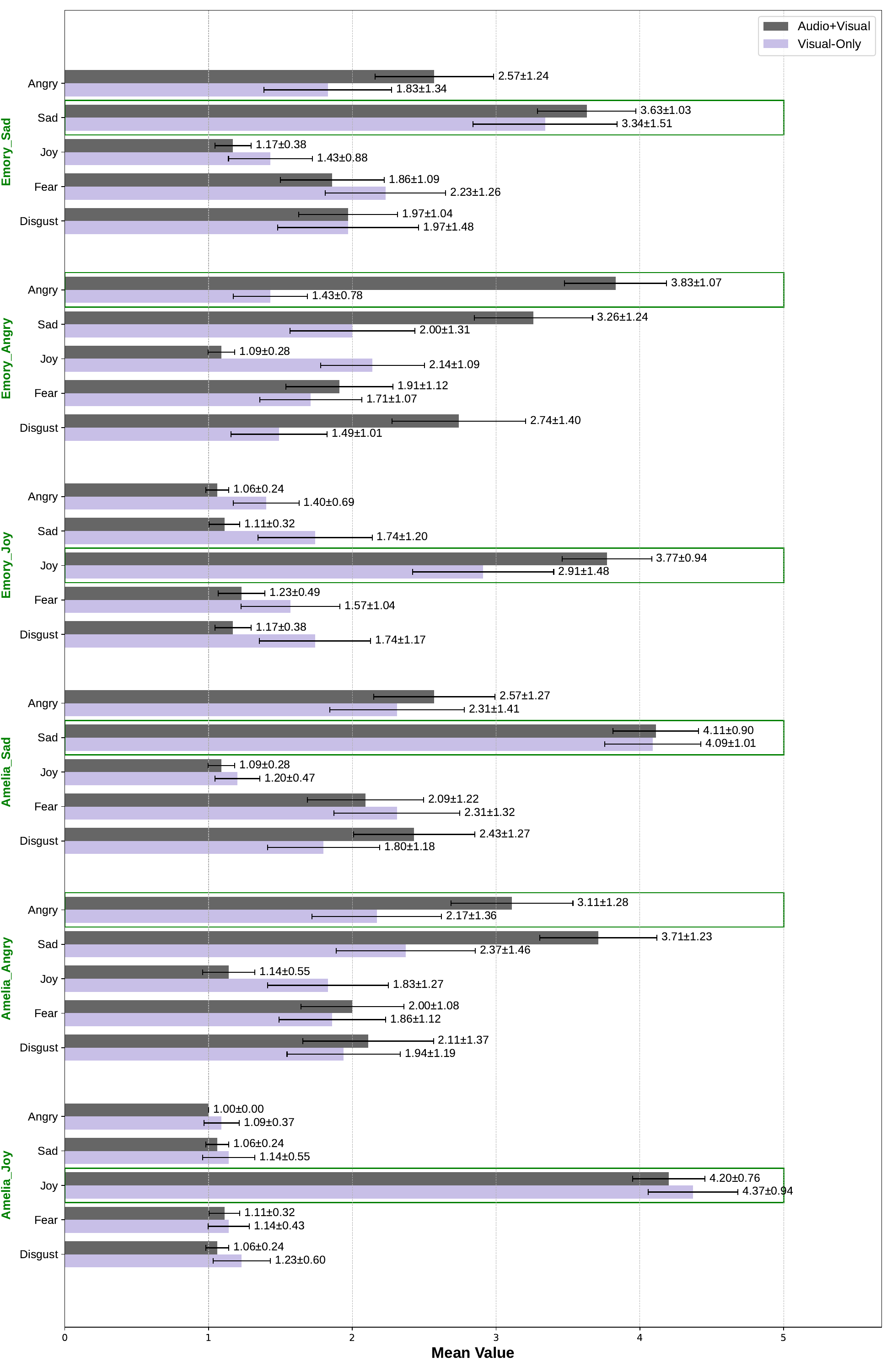}
\caption{Mean ratings ($M \pm SD$) of emotional expressiveness across five emotions (Angry, Sad, Joy, Fear, Disgust) for the avatars \textit{Emory} and \textit{Amelia}, under both audio+visual and visual-only conditions. Bars represent group means; values next to each bar report the mean and standard deviation, while horizontal error bars indicate 95\% confidence intervals (CI). Each video clip is labeled on the left, with color-coded bars distinguishing the experimental conditions. Green-highlighted boxes indicate target emotions critical for the study’s recognition task.
    }
\label{fig:plot}
\end{figure}

\subsubsection{Statistical Validation and Model Robustness}

To evaluate the reliability and robustness of the statistical modeling framework, we performed an extensive set of diagnostic procedures, model comparisons, and fit assessments across all three primary response categories. These included emotion recognition, perceived facial realism, and post-exposure ratings of realism contributors and empathic engagement.

For the emotion recognition analysis, model selection was guided by information-theoretic criteria, specifically the Akaike Information Criterion (AIC) and the Bayesian Information Criterion (BIC), along with likelihood ratio tests (LRTs) to assess violations of the proportional-odds (PO) assumption. As shown in Table~\ref{tab:val_em}, relaxing the Proportional Odds (PO) constraint significantly improved model fit ($p < 0.001$), justifying the use of more flexible specifications in the CLMMs. To ensure sufficient power, an a priori simulation-based power analysis was conducted, confirming that a sample size of $N = 70$ was adequate to detect medium-sized effects with a power of at least 0.80.

Similarly, Table~\ref{tab:val_fe_1} summarizes model comparisons for perceived facial expression realism. Again, significant improvements were observed when PO constraints were relaxed (e.g., LRT = 33.35, df = 15, $p = 0.004$), indicating that ordinal responses were not evenly distributed across thresholds. The best-performing CLMMs, detailed in Table~\ref{tab:val_fe_2}, achieved a conditional $R^2$ of 0.640, which indicates strong explanatory power when both fixed and random effects were considered. The marginal $R^2$ of 0.099 reflects the variance attributable to fixed effects alone. Residual diagnostics revealed an RMSE of 3.176 and residual standard deviation ($\sigma = 3.238$), confirming acceptable levels of predictive error and model stability.

In addition to these analyses, we assessed model robustness for the third response category, which included perceived realism across contributing aspects (e.g., facial expressions, visual appearance, voice tone, dialogue content) and overall empathy toward the avatars. Internal consistency for the multi-item realism contributors scale was found to be marginal, with a Cronbach’s alpha of $\alpha = 0.652$, suggesting moderate internal reliability. Furthermore, a logistic regression model predicting empathy responses yielded a Nagelkerke $R^2$ of just 0.011, indicating minimal variance explained. These findings suggest high individual variability and underscore the challenge of modeling subjective constructs like empathic connection using fixed predictors alone.

Across all three domains, model coefficients were exponentiated to derive interpretable odds ratios (OR) and visualized using forest plots. Where applicable, multiple comparisons were corrected using Holm’s method. To contextualize statistical findings, open-ended responses were triangulated with quantitative patterns, particularly regarding visual fidelity, emotional clarity, and audiovisual congruence. These effects were especially pronounced in high-arousal conditions, such as anger, where audiovisual alignment proved critical for accurate emotion recognition.

Taken together, these results validate the overall modeling approach and underscore its suitability for capturing the complex interplay between avatar design, experimental conditions, and emotional perception. The convergence of statistical rigor, diagnostic confirmation, and qualitative insight provides a robust foundation for interpreting user responses to emotionally expressive virtual avatars.

\begin{table}[H]
\centering
\caption{Model comparison for emotion recognition accuracy. Relaxing the PO assumption significantly improved model fit, as confirmed by the likelihood ratio test ($p < .001$).}
\label{tab:val_em}
\begin{tabular}{lccccl}
\hline
\multicolumn{1}{c}{\textbf{Df}} & \textbf{logLik} & \textbf{AIC} & \textbf{LRT} & \textbf{Pr(\textgreater{}Chisq)} & \multicolumn{1}{c}{\textbf{no.par}} \\\hline
6                               & -2521.47       & 5076.95     & 30.57       & 0.00                            & 17                                  \\ \hline
\end{tabular}
\end{table}

\begin{table}[H]
\centering
\caption{Model comparison for perceived facial expression realism. Relaxing the PO assumption improved model fit significantly across multiple model specifications.}
\label{tab:val_fe_1}
\begin{tabular}{cccccc}
\hline
\multicolumn{1}{l}{\textbf{Df}} & \multicolumn{1}{l}{\textbf{logLik}} & \multicolumn{1}{l}{\textbf{AIC}} & \multicolumn{1}{l}{\textbf{LRT}} & \multicolumn{1}{l}{\textbf{Pr(\textgreater{}Chi)}} \\ \hline
3.00                            & -382.84                            & 937.69                          & 9.34                            & 0.02                                              \\
6.00                            & -380.78                            & 939.57                          & 13.46                           & 0.03                                              \\
15.00                           & -370.84                            & 937.69                          & 33.34                           & 0.004                                              \\ \hline
\end{tabular}
\end{table}

\begin{table}[H]
\centering
\caption{Fit statistics for the best-performing CLMM in the facial realism analysis. Conditional and marginal $R^2$ values indicate explained variance; RMSE and $\sigma$ reflect prediction error and residual variability.}
\label{tab:val_fe_2}
\begin{tabular}{llllll}
\hline
\textbf{AIC}                 & \textbf{BIC}                 & \textbf{R2\_conditional}  & \textbf{R2\_marginal}     & \textbf{RMSE}             & \textbf{\boldmath$\sigma$}           \\ \hline
\multicolumn{1}{c}{1035.65} & \multicolumn{1}{c}{1100.30} & \multicolumn{1}{c}{0.64} & \multicolumn{1}{c}{0.09} & \multicolumn{1}{c}{3.17} & \multicolumn{1}{c}{3.23} \\ \hline
\end{tabular}
\end{table}

\subsubsection{Emotion Recognition} \label{sec:emotion}

\begin{figure}[H]
\centering
\includegraphics[scale=0.4, clip]{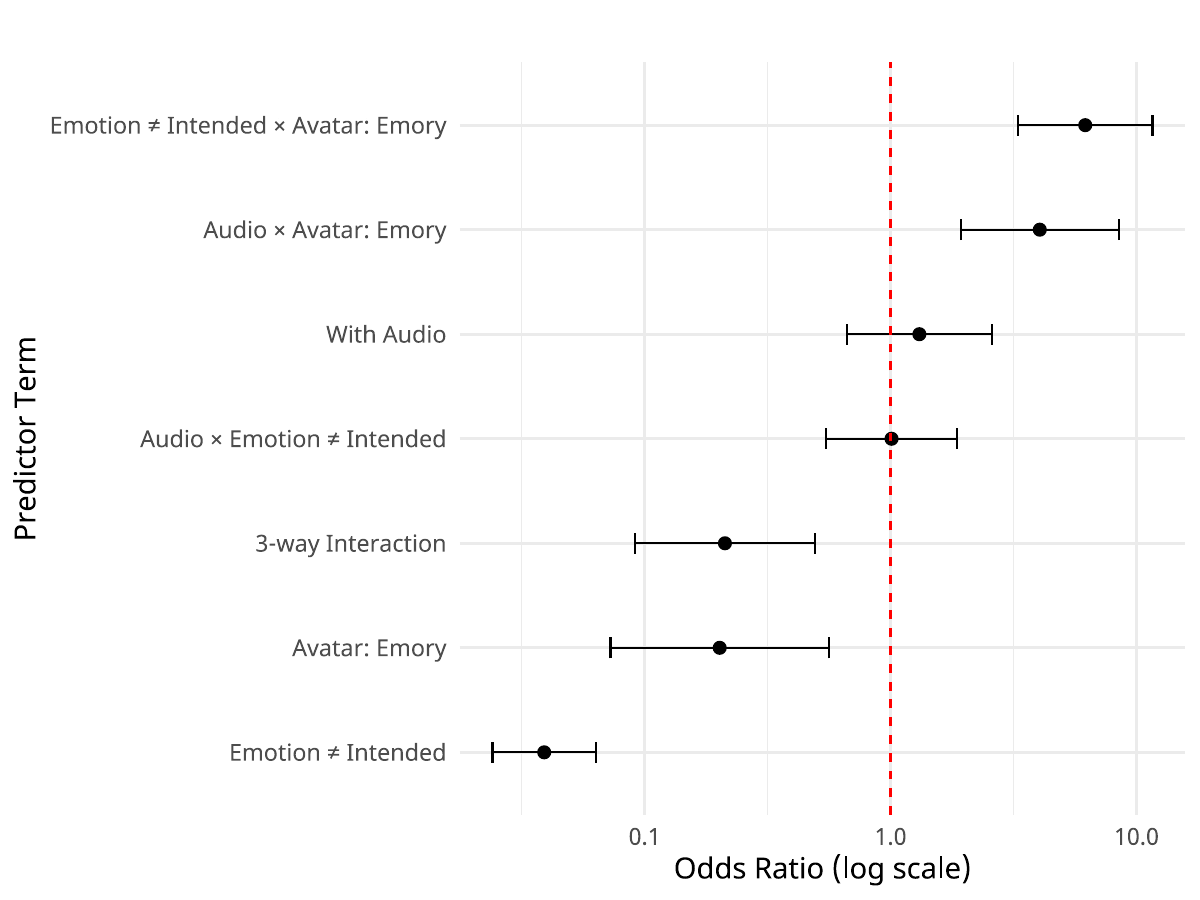}
\caption{Forest plot showing OR and 95\% CI for predictors of emotion recognition accuracy. The dashed vertical line indicates no effect (OR = 1). Statistically significant predictors are clearly indicated. \textit{Reference levels:} Avatar = Amelia, Condition = visual-only, Emotion = intended.}
\label{fig:forest_em}
\end{figure}

To initiate our analysis of emotion recognition, we present the results of the ordinal mixed-effects model (Table~\ref{tab:emotion_results}), supported by two key visualizations: a forest plot illustrating the influence of model predictors (Figure~\ref{fig:forest_em}) and a bar graph comparing perceived emotional expressiveness across experimental conditions and avatars identity (Figure~\ref{fig:plot}). Together, these results offer a comprehensive overview of participants’ accuracy in identifying the target emotions conveyed by the avatars.

Figure~\ref{fig:plot} shows that \textit{sadness} and \textit{joy} were recognized with high accuracy across both conditions. Sadness recognition remained stable for both avatars (Amelia: $M = 4.11$ vs. $M = 4.09$; Emory: $M = 3.63$ vs. $M = 3.34$), consistent with its strong reliance on static visual cues such as brow depression (AU1+4) and lip-corner droop (AU15) as defined in FACS~\cite{ekman1978facial}.

In contrast, recognition of joy varied by avatar identity and condition. Amelia’s joy remained consistently high, slightly improving in the visual-only condition ($M = 4.37$ vs. $M = 4.20$). Emory’s joy ratings, however, dropped 22\% in visual-only condition ($M = 3.77$ to $M = 2.91$). This disparity may reflect both visual ambiguity and a potential methodological limitation, namely the use of two different TTS applications, which may have introduced variations in emotional prosody, and the use of adult female voices for both avatars, which might not have fully matched their child-like appearance. Emory’s voice, generated with OpenAI’s TTS, exhibited strong emotional expressiveness, particularly in terms of pitch variation and intensity. However, this high affective prosody may not have been fully matched by the corresponding facial expressions, which might have contributed to a perceived mismatch between voice and visuals. In contrast, Amelia’s voice, synthesized using NaturalReaders, seemed to align more closely with the visual and emotional tone of the avatar, although this could also be influenced by other factors such as facial morphology or expression timing. Nevertheless, we cannot attribute the entire effect to the voice alone, as the study also included visual-only conditions in which Amelia’s avatar consistently performed better in terms of emotion recognition and perceived realism. This suggests that additional factors, such as facial morphology and expression clarity, also played a substantial role. Amelia’s advantage may be partly explained by her softer facial features and the presence of clear Duchenne markers (e.g., AU6+12), which enhanced visual clarity even in the absence of audio. In contrast, Emory’s more angular facial structure may have contributed to perceptual ambiguity, with features such as brow tension being more easily misinterpreted as negative affect. 

\begin{table}[H]
\centering

\caption{Ordinal mixed-effects model results predicting emotion recognition accuracy. Odds ratios (OR), 95\% confidence intervals (CI), and p-values are reported. Significant effects ($p < .05$) are highlighted. \textit{Reference levels:} Avatar = Amelia, Condition = visual-only, Emotion = intended. All model terms are interpreted as contrasts relative to these reference levels.}
\label{tab:emotion_results}
\begin{tabular}{lcccc}
\hline
\multicolumn{1}{c}{\textbf{term}}             & \textbf{p.value}              & \textbf{OR} & \textbf{OR\_low} & \textbf{OR\_high} \\ \hline
Emotion $\neq$ Intended                   & \cellcolor[HTML]{BFBFBF}0.000 & 0.039       & 0.024            & 0.063             \\
Audio+Visual                         & 0.437                         & 1.310       & 0.663            & 2.586             \\
Avatar: Emory                        & \cellcolor[HTML]{BFBFBF}0.002 & 0.202       & 0.073            & 0.562             \\
Audio+Visual × Emotion $\neq$   Intended         & 0.976                         & 1.010       & 0.547            & 1.864             \\
Emotion $\neq$ Intended   × Avatar: Emory & \cellcolor[HTML]{BFBFBF}0.000 & 6.191       & 3.299            & 11.618            \\
Audio+Visual × Avatar:   Emory              & \cellcolor[HTML]{BFBFBF}0.000 & 4.045       & 1.931            & 8.472             \\
3-way Interaction                    & \cellcolor[HTML]{BFBFBF}0.000 & 0.212       & 0.091            & 0.493             \\ \hline

\end{tabular}
\end{table}

These findings reinforce the idea that while sadness can be conveyed effectively through visual cues alone, accurate recognition of emotions like joy often depends on audiovisual congruence, particularly for avatars with less prototypical or more ambiguous facial expressions~\cite{pei2024affective}.

Among all emotions, anger recognition was the most negatively affected in the visual-only condition. Ratings for Emory fell from $M = 3.83$ to $M = 1.43$, and for Amelia from $M = 3.11$ to $M = 2.17$. Silent anger was often misinterpreted as sadness, reflecting the high reliance of anger on auditory cues such as spectral roughness and vocal intensity~\cite{gao2020fmri, marian2021cross}. Visual signals alone, particularly AU4, were insufficient for conveying high-arousal affect.

Even in the audio+visual condition, Amelia’s anger ($M = 3.11$) was frequently overshadowed by sadness ($M = 3.71$), suggesting perceptual interference from her softer features and higher-pitched voice. Emory’s more angular structure and pronounced brow tension better conveyed anger ($M = 3.83$ vs. $M = 3.26$ for sadness), underscoring the role of morphology in affect salience. This may have been further influenced by his highly emotive voice, which could have reinforced the expression of anger by providing strong auditory cues.

The statistical analysis (Table~\ref{tab:emotion_results}, Figure~\ref{fig:forest_em}) confirmed a significant three-way interaction among experimental conditions (audio+visual vs. visual-only), avatar identity, and emotion congruency ($p < .001$, OR = 0.212). This finding highlights the complex and interdependent nature of emotion recognition accuracy, suggesting that participants’ ability to correctly interpret emotional expressions was jointly shaped by the presentation mode, the specific avatar shown, and whether the expressed emotion aligned with the intended target.

In addition, the model identified a highly significant main effect for emotional incongruency. When the emotion recognized by the participant did not match the intended expression, the odds of accurate recognition dropped dramatically ($p < .001$, OR = 0.039, 95\% CI [0.024, 0.063]). This indicates that congruency between intended and perceived emotion was a critical factor in successful emotion interpretation.

Notably, the main effect of audio+visual condition alone was not statistically significant ($p = .437$, OR = 1.310), suggesting that the presence of auditory cues did not independently enhance recognition accuracy across conditions. Rather, the influence of audio appeared to depend on interactions with other variables, particularly avatar identity.

Avatar identity emerged as a significant factor in recognition performance. Specifically, participants were substantially less accurate in identifying emotions displayed by the avatar \textit{Emory} compared to \textit{Amelia} ($p = .002$, OR = 0.202, 95\% CI [0.073, 0.562]). Moreover, a significant interaction between experimental condition and avatar identity (\textit{Audio+Visual $\times$ Avatar: Emory}; $p < .001$, OR = 4.045, 95\% CI [1.931, 8.472]) indicated that the effect of audio cues seemed to vary depending on other factors—especially which avatar was shown.

Importantly, the interaction between incongruent emotional expressions and avatar identity (\textit{Emotion $\neq$ Intended $\times$ Avatar: Emory}; $p < .001$, OR = 6.191, 95\% CI [3.299, 11.618]) revealed that the negative impact of mismatched emotional cues was especially pronounced for Emory. This suggests that when Emory's facial expression did not match the target emotion, participants were particularly prone to misinterpretation, underscoring both the sensitivity of emotion recognition to the alignment between vocal and facial emotional cues, and the role of avatar-specific morphology in modulating perceptual accuracy.

One plausible explanation for Amelia’s superior emotion recognition lies in her softer facial morphology, which aligns more closely with users’ expectations of emotional authenticity~\cite{demeure2011believability}. Her rounded features and gentle brow dynamics enhanced the perceived realism of low-arousal emotions like sadness and joy. In contrast, Emory’s angular face and more pronounced brow movements made high-arousal expressions like anger more salient~\cite{marian2021cross, gao2020fmri}. However, beyond facial structure, the alignment between voice characteristics and facial appearance may also have played a role. Amelia’s relatively soft and emotionally neutral voice appeared to match her visual design, potentially reinforcing the perception of coherence. In contrast, Emory’s highly expressive voice may have occasionally clashed with his facial expressions, leading to moments of perceptual mismatch and ambiguity. This highlights a key design trade-off: while subtle facial geometry can enhance believability for certain emotions, it may reduce clarity for others, especially when not fully synchronized with vocal affect.

\subsubsection{Perceived Facial Expressions Realism} \label{sec:fe}

To investigate the perceived realism of facial expressions, we analyzed participants' ratings across two experimental conditions (\textit{audio+visual} and \textit{visual-only}) and two avatar identities.

\begin{figure}[H]
\centering
\includegraphics[scale=0.25]{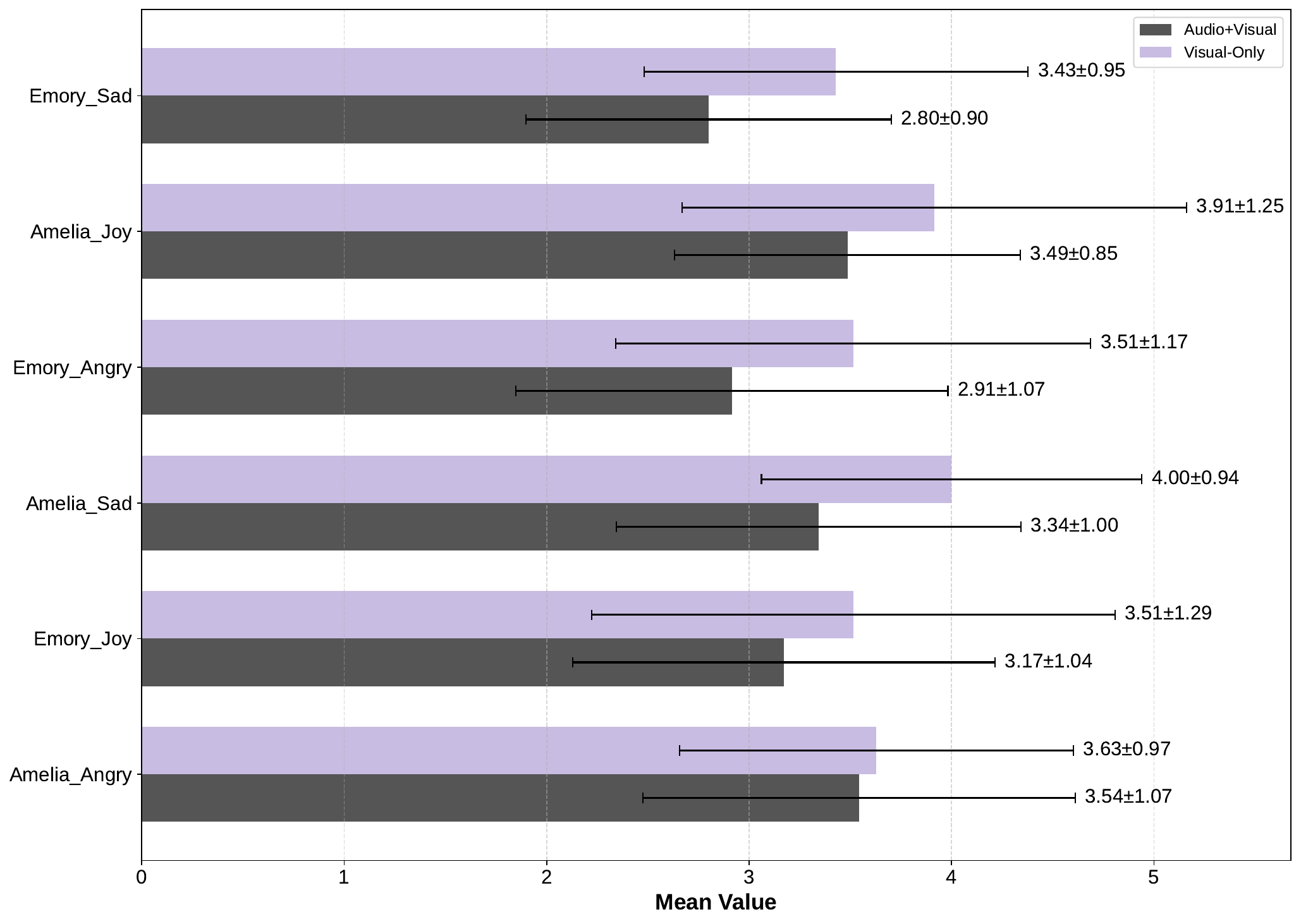}
\caption{Mean ratings ($M \pm SD$) perceived Facial Expressions realism for each video clips under both audio+visual and visual-only conditions. }
\label{fig:fe}
\end{figure}

\begin{figure}[H]
\centering
\includegraphics[scale=0.4, clip]{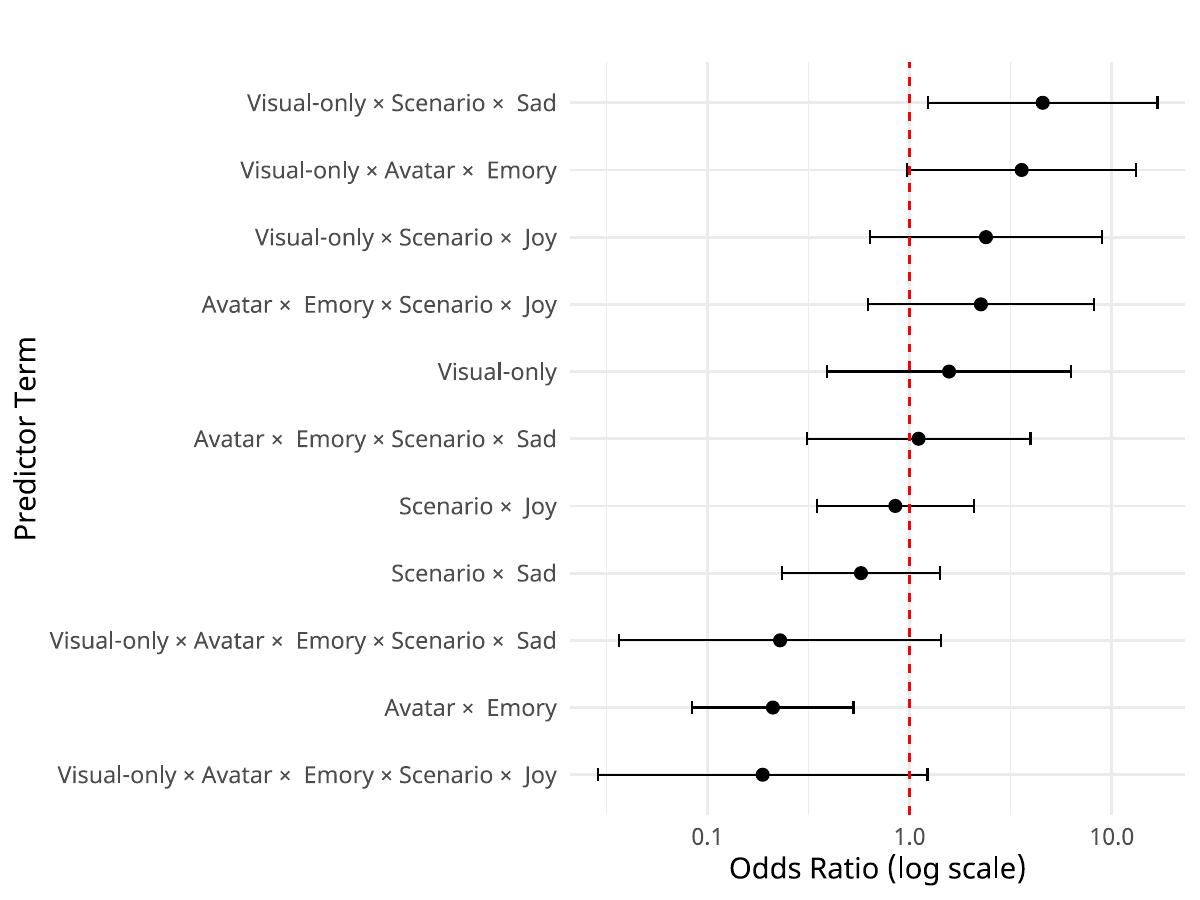}
\caption{Forest plot showing OR with 95\% CI for the model predictors of perceived realism of facial expressions. The vertical dashed line indicates no effect (OR = 1). Statistically significant predictors are clearly indicated. \textit{Reference levels:} Avatar = Amelia, Condition = Audio+Visual, Scenario = Angry. }
\label{fig:forest_fe}
\end{figure}

As shown in Figure~\ref{fig:fe}, facial expressions were consistently rated as more realistic in the \textit{visual-only} condition across all scenarios. This trend was further supported by the ordinal mixed-effects model (Table~\ref{tab:fe}; Figure~\ref{fig:forest_fe}). While the main effect of experimental condition was not statistically significant ($p = .526$), a significant interaction emerged between the visual-only condition and sad emotional content ($p = .023$, OR = 4.554, 95\% CI [1.232, 16.829]). This indicates that without audio, low-energy emotions like sadness seemed more realistic to viewers.

This enhancement likely stems from audiovisual interference. In the audio+visual condition, even slight mismatches in lip-sync or emotional prosody can disrupt the illusion of authenticity. The visual-only condition eliminates such dissonance, allowing facial cues to be evaluated in isolation, which may reduce perceptual scrutiny and heighten tolerance for animation artifacts.

A strong main effect of avatar identity was also observed: \textit{Emory} was rated significantly less realistic than \textit{Amelia} ($p = .001$, OR = 0.211, 95\% CI [0.084, 0.528]), regardless of experimental condition or emotion. As previously concluded in Section~\ref{sec:emotion}, Emory’s angular features and more rigid expressions may have clashed with participants’ implicit expectations of emotional expressiveness, whereas Amelia’s softer, more rounded features aligned better with perceived emotional authenticity. In addition, a good match between the voice and the avatar’s face likely made the avatar feel more realistic. Amelia’s voice was perceived as more consistent with her visual design, while Emory’s highly expressive vocal delivery may have introduced moments of mismatch when not fully supported by corresponding facial expressions.

\begin{table}[H]
\centering
\scriptsize
\caption{Results from the ordinal mixed-effects model predicting perceived realism of avatars' facial expressions. Odds ratios (OR), 95\% confidence intervals (CI), and p-values are presented. Significant effects ($p < .05$) are highlighted. \textit{Reference levels:} Avatar = Amelia, Condition = Audio+Visual, Scenario = Angry. All contrasts are interpreted relative to these reference levels.}

\label{tab:fe}
\begin{tabular}{lcccc}
\hline
\textbf{term}                                          & \multicolumn{1}{l}{\textbf{p.value}}   & \multicolumn{1}{l}{\textbf{OR}} & \multicolumn{1}{l}{\textbf{OR.lower}} & \multicolumn{1}{l}{\textbf{OR.upper}} \\ \hline
Visual-only                                   & 0.526                         & 1.568                  & 0.391                        & 6.294                        \\
Avatar: Emory                                 & \cellcolor[HTML]{BFBFBF}0.001 & 0.211                  & 0.084                        & 0.528                        \\
Scenario: Joy                                 & 0.722                         & 0.850                  & 0.348                        & 2.077                        \\
Scenario: Sad                                 & 0.228                         & 0.575                  & 0.234                        & 1.413                        \\
Visual-only × Avatar: Emory                   & 0.055                         & 3.584                  & 0.972                        & 13.208                       \\
Visual-only × Scenario: Joy                   & 0.198                         & 2.387                  & 0.635                        & 8.981                        \\
Visual-only × Scenario: Sad                   & \cellcolor[HTML]{BFBFBF}0.023 & 4.554                  & 1.232                        & 16.829                       \\
Avatar: Emory × Scenario: Joy                 & 0.216                         & 2.254                  & 0.622                        & 8.165                        \\
Avatar: Emory × Scenario: Sad                 & 0.875                         & 1.108                  & 0.310                        & 3.963                        \\
Visual-only × Avatar: Emory ×   Scenario: Joy & 0.081                         & 0.188                  & 0.029                        & 1.225                        \\
Visual-only × Avatar: Emory ×   Scenario: Sad & 0.115                         & 0.229                  & 0.037                        & 1.433                        \\ \hline
\end{tabular}
\end{table}

Although some interaction effects (e.g., Visual-only $\times$ Avatar: Emory, $p = .055$) did not reach statistical significance, the directional trend suggests meaningful interplay between avatar morphology and presentation condition, which merits further investigation.

In summary, low-arousal emotions like sadness appear more effectively conveyed through visuals alone, as unsynchronized audio can introduce perceptual dissonance that reduces realism. These findings underscore the importance of condition-sensitive avatar design, especially in emotionally nuanced applications such as child-simulation training, where voice-age congruence and expressive fidelity are crucial to achieving believable interaction.

\begin{figure}[H]
\centering
\includegraphics[scale=0.28]{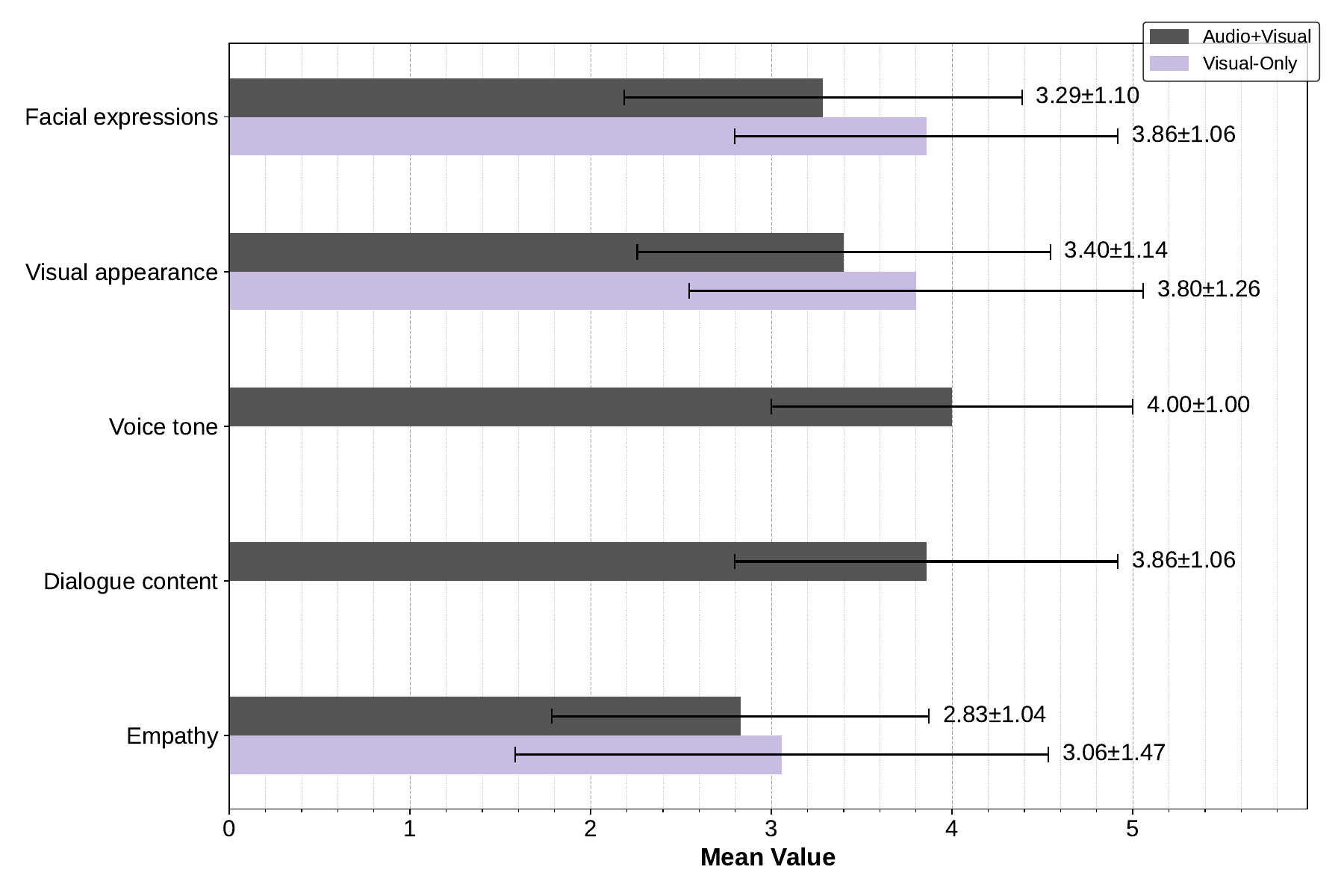}
\caption{Mean ratings ($M \pm SD$) of perceived realism across different aspects (facial expressions, visual appearance, voice tone, dialogue content) and empathy, compared between audio+visual and visual-only conditions. }
\label{fig:plot_section2}
\end{figure}

% \usepackage{multirow}
\begin{table}[H]
\centering
\caption{Results of mixed-effects cumulative-logit analyses comparing perceived realism and empathy between audio+visual and visual-only conditions. Significant results ($p < .05$) are highlighted.}
\label{tab:section2}
\begin{tabular}{llcccc}
\hline
                         & \textbf{term}                                          & \multicolumn{1}{l}{\textbf{p.value}} & \multicolumn{1}{l}{\textbf{OR}} & \multicolumn{1}{l}{\textbf{OR\_low}} & \multicolumn{1}{l}{\textbf{OR\_high}} \\ \hline
\multirow{3}{*}{\textbf{Realism}} & Audio+Visual vs. Visual-Only                  & \multicolumn{1}{c}{\cellcolor[HTML]{BFBFBF}{\color[HTML]{000000} 0.038}}   & 0.258                  & 0.072                       & 0.925                        \\
                         & Visual Appearance vs. Facial Expressions               & 0.765                        & 0.865                  & 0.333                       & 2.246                        \\
                         & Interaction: Audio+Visual × Visual Appearance & 0.587                        & 1.434                  & 0.391                       & 5.264                        \\ \hline
\textbf{Empathy}                  & Audio+Visual vs. Visual-Only                  & \multicolumn{1}{c}{0.397}   & 0.693                  & 0.295                       & 1.614                        \\ \hline
\end{tabular}
\end{table}

 \subsubsection{Perceived Realism and Empathy}

 The results summarized in Table~\ref{tab:section2} and Figure~\ref{fig:plot_section2} illustrate how multimodal cues (audio and visual) influenced participants’ perception of realism and empathy towards emotionally expressive avatars.

In the audio+visual condition, auditory elements played a major role in shaping perceptions. Voice tone emerged as the strongest contributor, with an average realism rating of $M = 4.00$, followed by dialogue content ($M = 3.86$). However, in the visual-only condition, visual components, particularly facial expression realism and visual appearance, received higher ratings ($M = 3.86$ and $M = 3.80$, respectively) compared to the audio+visual condition ($M = 3.29$ and $M = 3.40$, respectively). This difference may partly reflect the fact that participants in the visual-only condition evaluated fewer realism contributors, with their attention concentrated on the available visual aspects. In contrast, participants in the audio+visual condition distributed their judgments across four factors, including voice tone and dialogue content, which may have diluted individual ratings for visual features.

A mixed-effects cumulative-logit analysis confirmed these findings. Including audio led to a significant 74\% reduction in the likelihood of higher realism ratings (OR = 0.26, 95\% CI [0.07, 0.93], $p = .038$). This suggests that audio cues may introduce inconsistencies, such as minor lip-sync issues or mismatches between tone and visual emotion, which reduce perceived authenticity. No significant differences were found between the two realism items (``Facial Expressions'' vs. ``Visual Appearance,'' $p = .765$), nor was the interaction between condition and realism item significant ($p = .587$), indicating that audio affected both visual aspects similarly.

Empathy ratings were not significantly influenced by the presence of audio. Although slightly higher empathy was reported in the visual-only condition ($M = 3.06$) than in the audio+visual condition ($M = 2.83$), the difference was not statistically significant (OR = 0.69, 95\% CI [0.30, 1.61], $p = .397$). This suggests a possible trend toward enhanced emotional connection with visual-only stimuli, though not conclusive.

Collectively, these findings highlight a critical trade-off in avatar design: while audio cues enhance realism for high-arousal emotions, subtle audiovisual mismatches can negatively impact perceived authenticity. Under the conditions examined, removing audio increased visual credibility without substantially affecting emotional engagement.

\subsection{Qualitative Feedback}

Participants provided reflective qualitative feedback on what stood out the most about the avatars. These reflections revealed distinct and important observations regarding emotional expressiveness, realism, and immersion, providing directions for enhancing avatar technologies in interactive simulations. Two independent coders analyzed the responses. Discrepancies in coding were discussed and resolved collaboratively to ensure consistency and analytical rigor.

\subsubsection{Audio and Visual Conditions}

In the audio+visual condition, participants' qualitative feedback highlighted both the strengths and limitations of the avatars' emotional expressiveness. On the positive side, many participants recognized and appreciated the avatars' capacity to convey emotions dynamically, noting that \textit{"They were all expressing emotions"}  and that they \textit{"expressed their feelings through facials"}. The quality of the synthesized voice was frequently mentioned as a significant contributor to engagement and realism, with comments such as \textit{"The tone of voice was the most convincing"}  and \textit{"The dialogue content... did not sound like an avatar"}, suggesting successful prosody design and auditory fidelity. Some participants also felt a sense of connection, stating, \textit{"You can feel them though facial expressions"}. The visual appearance was also praised, with feedback indicating that \textit{"Their visual appearances were very good and looked human like"}  and \textit{"The faces were well done"}. 

However, a recurring theme in the feedback was the presence of multimodal incongruities. Participants frequently observed a mismatch between the auditory and visual channels, as captured by comments like \textit{"The facial expression does not match what the avatars are saying"}. Issues with lip synchronization were also noted as detrimental to realism, with feedback stating, \textit{"When they speak, their lips do not correspond to the dialogue"}  and \textit{"The mouth moves but it doesn't seam connected to the rest of the face"}. These observations align with the quantitative findings regarding reduced perceived facial realism when audio was present (Table \ref{tab:fe}), suggesting that the inclusion of voice amplified scrutiny of temporal and spatial alignment. 

Participants also reported experiencing uncanny valley effects, describing the avatars with phrases such as \textit{"weird mix of looking real but still obviously fake"}  and comparing their movement to \textit{"NPC characters in video games"}  or \textit{"a rigid model in video games"}. Some felt the facial expressions were not consistently on point  or that the faces \textit{"didnt move all that much"}. Negative emotional responses were also reported by some participants, with comments like \textit{"They all looked very scary"}  and \textit{"I feel uncomfortable and uneasy after this"}. 

Collectively, the feedback from the audio+visual condition reveals a tension between the successful implementation of some aspects, such as voice quality and general emotional expression, and persistent challenges related to audiovisual synchronization and the subtlety and naturalness of facial movements. These points underscore the complexity of creating truly believable and comfortable AI-generated avatars.

\subsubsection{Visual-Only Conditions}

In the visual-only condition, where avatars were presented without synthesized speech, participant feedback predominantly focused on visual realism and the effectiveness of facial cues in conveying emotion. Positive comments highlighted the avatars' ability to express emotions through isolated facial dynamics, with participants noting that \textit{"Facial expressions stood out the most from the avatars"} and that the avatars showed emotion \textit{"like real human"}. Some feedback specifically praised the clarity of certain expressions, such as joy and sadness, with one participant observing a \textit{"genuine smile, often characterized by crinkles around the eyes"} for happiness and another emphasizing the effectiveness of \textit{"The emotional that was in the eyes of avatars... specially the emotion of sadness and disgust,"} adding \textit{"good job!"}. The visual design was also acknowledged positively, with comments like \textit{"They have human-like features"} and \textit{"Amelia and Emory looked the most realistic to me"}. Some participants reported feeling \textit{"Very connected with the avatar"} and that \textit{"You can feel them though facial expressions"} even in the absence of audio.

However, critiques were also present, centering on perceived exaggeration or inconsistency in the expressions. Participants reported that \textit{"facial expressions were quite exaggerated and did not look completely natural"} and that \textit{"They almost all seem to have the same facial expression if I did not look closely"}. Technical limitations in animation fidelity were also noted, with comments such as \textit{"I think they lack polygons to make them more expresive"} and comparisons to \textit{"a rigid model in video games"}. The absence of audio seemed to shift attention to subtle visual flaws, including repetitive mouth movements (\textit{"the movement of their mouth"}) and limited emotional range or gradation (\textit{"I would see a slight smile here and there but the avatar still kept a straight face"}). Some feedback also simply stated, \textit{"nothing"} stood out, or described the experience as \textit{"interesting but not very realistic"}.

Collectively, the feedback from the visual-only condition suggests that while removing audio can enhance the focus on and perceived realism of facial expressions by eliminating cross-modal interference, it also exposes limitations in the subtlety, naturalness, and fidelity of the facial animations themselves. These points underscore the need for further refinement in visual-only emotional expression to achieve greater believability.

\section{Discussion}

This study offers several important lessons regarding the perceptual and design challenges of building emotionally expressive child avatars. Rather than emphasizing the system’s technical capabilities, our findings point to deeper tensions in achieving believable emotional portrayals through multimodal integration. One key insight is that perceived realism depends not on modality fidelity in isolation, but on the coherence across modalities. Moreover, participant responses varied considerably, revealing the inherently subjective nature of emotional perception. What one viewer perceives as lifelike or relatable, another may find unsettling or unnatural. This variability underscores the importance of designing emotionally expressive avatars that can accommodate a range of user sensitivities. When auditory and visual channels are not well aligned, user experience suffers. In some cases, visual-only presentations were even rated as more realistic than audio-visual ones. Silencing the audio removes a major source of incongruence, but it can also weaken the communication of certain emotions, particularly those that involve high arousal.

The study highlights a recurring trade-off. Auditory cues are necessary for disambiguating intense emotional states such as anger, yet their inclusion risks drawing attention to imperfections in lip-sync or mismatched vocal tone. Synchronizing prosody with facial animation is not a small technical issue. It plays a key role in making avatars believable and emotionally clear. Participant feedback supports this, with several comments describing distracting or unsettling effects when lip movements did not match the dialogue or when the voice tone conflicted with the avatar’s appearance.

A second lesson concerns the role of avatar morphology in shaping emotional interpretation. The "Amelia" avatar, with softer facial features, was generally perceived as more believable in low-arousal contexts such as sadness or joy. In contrast, the more angular "Emory" avatar was more effective in conveying anger. These outcomes suggest that certain facial structures are better suited to subtle, empathic expressions, while others may more effectively convey direct, high-intensity emotions. Designers of virtual agents may benefit from selecting or customizing facial geometry based on the emotional demands of the specific use case.

Additionally, the study raises questions about the optimal configuration of animation parameters. Our use of the “exaggerated” mode in Audio2Face, intended to increase clarity, some
participants in the visual-only condition perceived the resulting facial movements as overly intense or unnatural. This suggests a potential trade-off between expressiveness and realism, especially in training contexts where exaggerated expressions may break immersion or reduce credibility. Exploring "automatic" or more customized settings could help address this issue.

%Taken together, these findings point to several design implications. First, when voice is included, investing in speech synthesis that is age-appropriate and emotionally congruent is essential. A poorly matched voice track can undermine even a well-animated avatar, while a congruent one can enhance weaker visuals. Second, in applications focused on empathy and subtle affect, such as child interview simulations, morphologies like Amelia’s may offer perceptual advantages. Finally, developers should consider providing fallback visual-only modes or adaptive modality controls, particularly in situations where audiovisual misalignment could hinder user trust or emotional engagement.

Overall, this work reinforces that perceived believability is not simply a function of system complexity or realism in individual channels, but rather the result of coherent integration across modalities and sensitivity to user perception. By focusing on these lessons, future systems can be better tailored to the emotional and communicative needs of users, especially in sensitive and high-stakes environments.

\subsection{Limitations and Future Work}

While this study offers several valuable insights into the design and perception of emotionally expressive child avatars, it also presents a number of limitations that inform directions for future research. The findings are based on two specific MetaHuman avatars, and user perceptions may differ with avatars of varying visual design, age representation, or artistic style. Moreover, although prior work supported the use of a young adult female voice for child avatars, this choice may have influenced participants’ perceptions of emotional authenticity and audiovisual coherence. Integrating high-quality, age-appropriate synthesized voices that align more naturally with the avatars' appearance and emotional tone remains a significant technical and perceptual challenge.

The emotional scope of the study was also limited to three basic emotions: joy, sadness, and anger. As a result, the system’s ability to portray a broader range of complex or subtle emotional states (such as fear, surprise, or confusion) remains untested. Additionally, participant recruitment via the Prolific platform, while offering demographic diversity, may not fully capture the perspectives of domain-specific end-users such as Child Protection Services (CPS) professionals. The controlled, pre-rendered nature of the video stimuli further limits ecological validity. Responses observed in dynamic, real-time interactions may differ substantially due to the role of responsiveness and timing in social perception.

These limitations point to several promising avenues for future work. Longitudinal studies are needed to explore how user perceptions evolve with repeated or sustained exposure to emotionally expressive avatars, potentially revealing learning effects and long-term engagement dynamics. Future research should also expand the emotional repertoire of avatars to better reflect the range of expressions relevant in sensitive contexts such as forensic interviews. Critically, upcoming evaluations should involve domain experts such as CPS or law enforcement professionals to assess the system’s usability, realism, and training efficacy in practical, high-stakes environments. Such studies will help refine both the technical and perceptual aspects of the system, ensuring its relevance and effectiveness in its intended application context.

\section{Conclusion}

This study highlights a central challenge in the development of emotionally expressive child avatars: effective multimodal integration remains difficult to achieve. While our system demonstrates the technical feasibility of real-time emotional animation using Unreal Engine 5 MetaHuman and NVIDIA Audio2Face, the results reveal that audiovisual misalignment can undermine perceived authenticity. Notably, participants often rated visual-only clips as more realistic than audio-visual ones, suggesting that incomplete or poorly matched modalities may detract from overall believability.

This is especially relevant in sensitive applications such as child forensic interview training, where avatars must evoke trust and empathy. Our findings show that high-arousal emotions like anger depend heavily on synchronized vocal and facial cues, while lower-arousal states like sadness and joy can be conveyed more effectively through visual expressions alone. Additionally, factors such as avatar morphology and voice-age congruence significantly shaped emotional perception.

Rather than pursuing realism within individual modalities, this study underscores the importance of holistic and context-sensitive design. Emotional credibility arises only when voice, appearance, and facial behavior are cohesively aligned. These insights offer practical guidance for improving the design of virtual agents in emotionally demanding settings. Ultimately, consistent integration across modalities, rather than fidelity in isolation, is essential for achieving believable human–AI interaction.

%\clearpage
%\bibliographystyle{IEEEtran}
%\bibliography{references}


%\clearpage
\appendix
\section*{Appendix: Supplementary Material}
\addcontentsline{toc}{section}{Appendix: Supplementary Materials}

This appendix includes supplementary statistical tables (Table~\ref{tab:main_t-test} and Table~\ref{tab:t-test-avatar}) that report model comparisons and detailed regression results supporting the analyses presented in Section~\ref{sec:result}.

% \usepackage{multirow}
\begin{table}[H]
\centering
\scriptsize
\caption{Detailed results of independent-sample t-tests (including Welch's corrections where needed) comparing visual-only and audio+visual conditions across emotional expressions and avatars. Equality of variances was assessed using Levene’s test. Cohen’s d reflects effect size magnitude. FDR-corrected p-values account for multiple comparisons. Statistically significant results (FDR-adjusted $p < .05$) are highlighted in gray.}
\label{tab:main_t-test}
\begin{tabular}{llccccc}
\hline
                                & \textbf{Variable} &  \textbf{\begin{tabular}[c]{@{}c@{}}Mean \\ visual-only\end{tabular}} & \textbf{\begin{tabular}[c]{@{}c@{}}Mean \\ audio+visual\end{tabular}} & \textbf{ p-value} & \textbf{Cohen's d} & \textbf{\begin{tabular}[c]{@{}c@{}}FDR-corrected \\ p-value\end{tabular}} \\ \hline
                                & Facial exp.       & 3.429                       & 2.800                        & 0.006                   & 0.680              & \cellcolor[HTML]{BFBFBF}0.021    \\
                                & Angry             & 1.829                       & 2.571                        & 0.019                   & -0.575             & 0.053                            \\
                                & Sad               & 3.343                       & 3.629                        & 0.360                   & -0.221             & 0.520                            \\
                                & Joy               & 1.429                       & 1.171                        & 0.119                   & 0.378              & 0.222                            \\
                                & Fear              & 2.229                       & 1.857                        & 0.192                   & 0.315              & 0.312                            \\
\multirow{-6}{*}{\rotatebox{90}{Emory\_Sad}}    & Disgust           & 1.971                       & 1.971                        & 1.000                   & 0.000              & 1.000                            \\ \hline
                                & Facial exp.       & 3.514                       & 2.914                        & 0.028                   & 0.535              & 0.074                            \\
                                & Angry             & 1.429                       & 3.829                        & 0.000                   & -2.565             & \cellcolor[HTML]{BFBFBF}0.000    \\
                                & Sad               & 2.000                       & 3.257                        & 0.000                   & -0.985             & \cellcolor[HTML]{BFBFBF}0.001    \\
                                & Joy               & 2.143                       & 1.086                        & 0.000                   & 1.329              & \cellcolor[HTML]{BFBFBF}0.000    \\
                                & Fear              & 1.714                       & 1.914                        & 0.448                   & -0.182             & 0.540                            \\
\multirow{-6}{*}{\rotatebox{90}{Emory\_Angry}}  & Disgust           & 1.486                       & 2.743                        & 0.000                   & -1.029             & \cellcolor[HTML]{BFBFBF}0.001    \\ \hline
                                & Facial exp.       & 3.514                       & 3.171                        & 0.226                   & 0.292              & 0.339                            \\
                                & Angry             & 1.400                       & 1.057                        & 0.008                   & 0.661              & \cellcolor[HTML]{BFBFBF}0.027    \\
                                & Sad               & 1.743                       & 1.114                        & 0.005                   & 0.717              & \cellcolor[HTML]{BFBFBF}0.021    \\
                                & Joy               & 2.914                       & 3.771                        & 0.005                   & -0.690             & \cellcolor[HTML]{BFBFBF}0.021    \\
                                & Fear              & 1.571                       & 1.229                        & 0.082                   & 0.423              & 0.177                            \\
\multirow{-6}{*}{\rotatebox{90}{Emory\_Joy}}    & Disgust           & 1.743                       & 1.171                        & 0.009                   & 0.656              & \cellcolor[HTML]{BFBFBF}0.027    \\ \hline
                                & Facial exp.       & 4.000                       & 3.343                        & 0.006                   & 0.678              & \cellcolor[HTML]{BFBFBF}0.021    \\
                                & Angry             & 2.314                       & 2.571                        & 0.425                   & -0.192             & 0.540                            \\
                                & Sad               & 4.086                       & 4.114                        & 0.901                   & -0.030             & 0.925                            \\
                                & Joy               & 1.200                       & 1.086                        & 0.224                   & 0.293              & 0.339                            \\
                                & Fear              & 2.314                       & 2.086                        & 0.455                   & 0.179              & 0.540                            \\
\multirow{-6}{*}{\rotatebox{90}{Amelia\_Sad}}   & Disgust           & 1.800                       & 2.429                        & 0.036                   & -0.513             & 0.081                            \\ \hline
                                & Facial exp.       & 3.629                       & 3.543                        & 0.726                   & 0.084              & 0.787                            \\
                                & Angry             & 2.171                       & 3.114                        & 0.004                   & -0.714             & \cellcolor[HTML]{BFBFBF}0.021    \\
                                & Sad               & 2.371                       & 3.714                        & 0.000                   & -0.997             & \cellcolor[HTML]{BFBFBF}0.001    \\
                                & Joy               & 1.829                       & 1.143                        & 0.005                   & 0.700              & \cellcolor[HTML]{BFBFBF}0.021    \\
                                & Fear              & 1.857                       & 2.000                        & 0.589                   & -0.130             & 0.656                            \\
\multirow{-6}{*}{\rotatebox{90}{Amelia\_Angry}} & Disgust           & 1.943                       & 2.114                        & 0.577                   & -0.134             & 0.656                            \\ \hline
                                & Facial exp.       & 3.914                       & 3.486                        & 0.098                   & 0.401              & 0.200                            \\
                                & Angry             & 1.086                       & 1.000                        & 0.179                   & 0.325              & 0.304                            \\
                                & Sad               & 1.143                       & 1.057                        & 0.400                   & 0.203              & 0.540                            \\
                                & Joy               & 4.371                       & 4.200                        & 0.405                   & 0.200              & 0.540                            \\
                                & Fear              & 1.143                       & 1.114                        & 0.754                   & 0.075              & 0.795                            \\
\multirow{-6}{*}{\rotatebox{90}{Amelia\_Joy}}   & Disgust           & 1.229                       & 1.057                        & 0.119                   & 0.377              & 0.222                            \\ \cline{2-7} 
\end{tabular}
\end{table}
% \usepackage{multirow}
\begin{table}[H]

\caption{Statistical comparison using paired-sample t-tests of emotional expression ratings between the Amelia and Emory avatars using paired-sample t-tests, under both Audio+Visual and Visual-Only conditions. Ratings include facial expressiveness and perceived intensity across five target emotions (Angry, Sad, Joy, Fear, Disgust). Cohen’s d reflects effect size. FDR-corrected p-values account for multiple comparisons. Statistically significant differences (FDR-adjusted $p < .05$) are highlighted in gray. For each significant result, the avatar with higher mean ratings is highlighted in green to indicate user preference.}
\centering
\scriptsize
\label{tab:t-test-avatar}
\begin{tabular}{clcccccccccc}
\hline
\multicolumn{1}{l}{}             &                   & \multicolumn{5}{c}{\textbf{Audio+Visual}}                                                                                                                                                                                                                                                                                          & \multicolumn{5}{c}{\textbf{Visual-Only}}                                                                                                                                                                                                                                                                                            \\ \hline
                                 & \textbf{Variable} & \multicolumn{1}{l}{\textbf{p-value}} & \multicolumn{1}{l}{\textbf{\begin{tabular}[c]{@{}l@{}}Mean\\ Amelia\end{tabular}}} & \multicolumn{1}{l}{\textbf{\begin{tabular}[c]{@{}l@{}}Mean\\ Emory\end{tabular}}} & \multicolumn{1}{l}{\textbf{Cohen's d}} & \textbf{\begin{tabular}[c]{@{}c@{}}FDR-corrected \\ p-value\end{tabular}} & \multicolumn{1}{l}{\textbf{p-value}} & \multicolumn{1}{l}{\textbf{\begin{tabular}[c]{@{}l@{}}Mean\\ Amelia\end{tabular}}} & \multicolumn{1}{l}{\textbf{\begin{tabular}[c]{@{}l@{}}Mean\\ Emory\end{tabular}}} & \multicolumn{1}{l}{\textbf{Cohen's d}} & \textbf{\begin{tabular}[c]{@{}c@{}}FDR-corrected\\   p-value\end{tabular}} \\
                                 & Facial exp.       & 0.001                                & 3.343                                                                              & 2.800                                                                             & 0.591                                  & \cellcolor[HTML]{BFBFBF}0.010                                             & 0.001                                & 4.000                                                                              & 3.429                                                                             & 0.646                                  & \cellcolor[HTML]{BFBFBF}0.005                                              \\
                                 & Angry             & 1.000                                & 2.571                                                                              & 2.571                                                                             & 0.000                                  & 1.000                                                                     & 0.088                                & 2.314                                                                              & 1.829                                                                             & 0.297                                  & 0.143                                                                      \\
                                 & Sad               & 0.020                                & 4.114                                                                              & 3.629                                                                             & 0.414                                  & \cellcolor[HTML]{BFBFBF}0.045                                             & 0.018                                & 4.086                                                                              & 3.343                                                                             & 0.419                                  & \cellcolor[HTML]{BFBFBF}0.041                                              \\
                                 & Joy               & 0.263                                & 1.086                                                                              & 1.171                                                                             & -0.192                                 & 0.338                                                                     & 0.118                                & 1.200                                                                              & 1.429                                                                             & -0.271                                 & 0.163                                                                      \\
                                 & Fear              & 0.244                                & 2.086                                                                              & 1.857                                                                             & 0.201                                  & 0.338                                                                     & 0.619                                & 2.314                                                                              & 2.229                                                                             & 0.085                                  & 0.619                                                                      \\
\multirow{-7}{*}{\textbf{\rotatebox{90}{Sad}}}   & Disgust           & 0.014                                & 2.429                                                                              & 1.971                                                                             & 0.440                                  & \cellcolor[HTML]{BFBFBF}0.041                                             & 0.280                                & 1.800                                                                              & 1.971                                                                             & -0.186                                 & 0.336                                                                      \\ \hline
                                 & Facial exp.       & 0.002                                & 3.543                                                                              & 2.914                                                                             & 0.578                                  & \cellcolor[HTML]{BFBFBF}0.010                                             & 0.579                                & 3.629                                                                              & 3.514                                                                             & 0.095                                  & 0.613                                                                      \\
                                 & Angry             & 0.002                                & 3.114                                                                              & 3.829                                                                             & -0.582                                 & \cellcolor[HTML]{BFBFBF}0.010                                             & 0.006                                & 2.171                                                                              & 1.429                                                                             & 0.495                                  & \cellcolor[HTML]{BFBFBF}0.016                                              \\
                                 & Sad               & 0.009                                & 3.714                                                                              & 3.257                                                                             & 0.466                                  & \cellcolor[HTML]{BFBFBF}0.041                                             & 0.113                                & 2.371                                                                              & 2.000                                                                             & 0.275                                  & 0.163                                                                      \\
                                 & Joy               & 0.571                                & 1.143                                                                              & 1.086                                                                             & 0.097                                  & 0.643                                                                     & 0.162                                & 1.829                                                                              & 2.143                                                                             & -0.242                                 & 0.208                                                                      \\
                                 & Fear              & 0.609                                & 2.000                                                                              & 1.914                                                                             & 0.087                                  & 0.644                                                                     & 0.483                                & 1.857                                                                              & 1.714                                                                             & 0.120                                  & 0.543                                                                      \\
\multirow{-6}{*}{\textbf{\rotatebox{90}{Angry}}} & Disgust           & 0.012                                & 2.114                                                                              & 2.743                                                                             & -0.451                                 & \cellcolor[HTML]{BFBFBF}0.041                                             & 0.004                                & 1.943                                                                              & 1.486                                                                             & 0.516                                  & \cellcolor[HTML]{BFBFBF}0.016                                              \\ \hline
                                 & Facial exp.       & 0.054                                & 3.486                                                                              & 3.171                                                                             & 0.337                                  & 0.108                                                                     & 0.041                                & 3.914                                                                              & 3.514                                                                             & 0.358                                  & 0.075                                                                      \\
                                 & Angry             & 0.160                                & 1.000                                                                              & 1.057                                                                             & -0.243                                 & 0.262                                                                     & 0.006                                & 1.086                                                                              & 1.400                                                                             & -0.498                                 & \cellcolor[HTML]{BFBFBF}0.016                                              \\
                                 & Sad               & 0.324                                & 1.057                                                                              & 1.114                                                                             & -0.169                                 & 0.389                                                                     & 0.002                                & 1.143                                                                              & 1.743                                                                             & -0.565                                 & \cellcolor[HTML]{BFBFBF}0.012                                              \\
                                 & Joy               & 0.020                                & 4.200                                                                              & 3.771                                                                             & 0.413                                  & \cellcolor[HTML]{BFBFBF}0.045                                             & 0.000                                & 4.371                                                                              & 2.914                                                                             & 1.025                                  & \cellcolor[HTML]{BFBFBF}0.000                                              \\
                                 & Fear              & 0.254                                & 1.114                                                                              & 1.229                                                                             & -0.196                                 & 0.338                                                                     & 0.026                                & 1.143                                                                              & 1.571                                                                             & -0.392                                 & 0.053                                                                      \\
\multirow{-6}{*}{\textbf{\rotatebox{90}{Joy}}}   & Disgust           & 0.103                                & 1.057                                                                              & 1.171                                                                             & -0.283                                 & 0.186                                                                     & 0.006                                & 1.229                                                                              & 1.743                                                                             & -0.495                                 & \cellcolor[HTML]{BFBFBF}0.016                                              \\ \hline
\end{tabular}
\end{table}

Between-condition evaluations in Table~\ref{tab:main_t-test} reveal a robust modality effect: when the audio track was muted, participants rated the avatars’ facial expressions as more realistic overall, with significantly higher ratings in the sad scenarios for both characters (Emory-Sad: $M_{\text{no-audio}} = 3.43$ vs $M_{\text{audio}} = 2.80$, $d = 0.68$, FDR-corrected $p = .021$; Amelia-Sad: $M_{\text{no-audio}} = 4.00$ vs $M_{\text{audio}} = 3.34$, $d = 0.67$, FDR-corrected $p = .021$). This baseline shift suggests that any mismatch between vocal prosody and facial timing, or the use of an ill-suited voice, draws attention and diminishes perceived realism.

Conversely, in scenarios depicting high-arousal emotions (Emory-Angry, Amelia-Angry, and Emory-Joy), the presence of audio improved emotion recognition and overall expressiveness. This pattern indicates that synchronized vocal cues can enhance the clarity of emotional intent, particularly when facial expressions alone provide insufficient affective information.

Table~\ref{tab:t-test-avatar} provides a direct comparison between the two avatars. Across both conditions, Amelia was consistently rated as more realistic, with higher accuracy in emotion identification for sadness and joy. Notably, in the audio-visual condition, Emory was rated more believable in the angry scenario, suggesting that his angular features may better support high-arousal affect when complemented by vocal intensity. In contrast, Amelia’s rounded, soft-featured face reliably conveyed emotional states across modalities, enabling consistent perception even without audio.

Taken together, these two tables reinforce our central claim: perceived believability depends critically on the interplay between audiovisual congruence and facial geometry.

%\clearpage
%\includepdf[pages=-]{files/supplementary.pdf}

\end{document}